\documentclass[sigconf]{acmart}
\usepackage{array}
\usepackage{balance}
\usepackage{appendix}
\usepackage{longtable}
\usepackage{xcolor}
\usepackage{listings}
\usepackage{graphicx}  % To scale the content
\usepackage{xspace}
\usepackage{subcaption}

\renewcommand\footnotetextcopyrightpermission[1]{}
\AtBeginDocument{%
  \providecommand\BibTeX{{%
    \normalfont B\kern-0.5em{\scshape i\kern-0.25em b}\kern-0.8em\TeX}}}

\begin{document}

\hyphenation{caus-ed}
% \newcommand{\myparagraph}[1]{\vspace{0.6\baselineskip}\noindent{\textbf{#1}}.~}

%%% adversarial IR
\newcommand{\auth}{publisher\xspace}
\newcommand{\auths}{publishers\xspace}
\newcommand{\SEO}{SEO\xspace}
\newcommand{\payoff}{utility\xspace}
\newcommand{\payoffs}{utilities\xspace}

%%% editing macros
\newcommand{\omt}[1]{}
\newcommand{\firstmention}[1]{{\bf #1}}

%%% basic quantities
\newcommand{\round}{l}
\newcommand{\order}{\pi}
\newcommand{\totalOrder}{TO}
\newcommand{\rankFunc}{r}
\newcommand{\numAuth}{n}
\newcommand{\query}{q}
\newcommand{\doc}{d}
\newcommand{\docSet}{D}
\newcommand{\initDocSet}{D^{0}}
\newcommand{\docinit}{d_{\rm init}}
\newcommand{\corpus}{{\cal D}}
\newcommand{\initTag}{{\rm init}}
\newcommand{\topRetGroup}{{\cal D}_{\initTag}}
\newcommand{\numRel}{n}
\newcommand{\arbTerm}{t}
\newcommand{\termsSet}{\mathcal{S}}

%%% general math
\newcommand{\set}[1]{\{#1\}}
\newcommand{\definedas}{\stackrel{def}{=}}
\newcommand{\rankEquiv}{\stackrel{rank}{=}}
\newcommand{\kron}[1]{\delta\hspace{-0.2111em}\left[#1\right]}
\newcommand{\kld}[2]{D\left(#1 \; \Big\vert\Big\vert \,\, #2\right)}
\newcommand{\kldSmall}[2]{D\left(#1 \; \vert\vert \,\, #2\right)}
\newcommand{\freq}[2]{{\rm tf}(#1 \in #2)}
\newcommand{\wordIndex}{j} 
\newcommand{\entropy}{H}
\newcommand{\crossEnt}{CE}
\newcommand{\ce}[2]{\crossEnt\left(#1 \; \Big\vert\Big\vert \,\, #2\right)}

%%% probability notation
\newcommand{\prob}{p}
\newcommand{\probhat}{\hat{\prob}}
\newcommand{\modelprob}{\hat{\prob}}
\newcommand{\lmprob}{\modelprob^{LM}}
\newcommand{\ilmprob}{\prob} %for decorating induced probs, if necessary
\newcommand{\ilmprobtil}{\tilde{\prob}}
\newcommand{\baseprob}{\prob}
\newcommand{\condArbP}[3]{\ensuremath{#1(#2 \vert #3)}}
\newcommand{\condP}[2]{\condArbP{\prob}{#1}{#2}}
\newcommand{\estCondP}[2]{\hat{\prob}(#1\vert#2)}
\newcommand{\condPhat}[2]{\hat{\prob}(#1|#2)}
\newcommand{\modelcondP}[2]{\condArbP{\modelprob}{#1}{#2}}
\newcommand{\basecondP}[2]{\condArbP{\baseprob}{#1}{#2}}
\newcommand{\lmcondP}[2]{\condArbP{\lmprob}{#1}{#2}}
\newcommand{\genprob}[2]{\inducedprob{\ilmprob}{#1}{#2}} %text string goes second
\newcommand{\MLE}{MLE\xspace}
\newcommand{\GEN}{GEN\xspace}

%%% tables
\newcommand{\weight}{$\Delta$MRR\xspace}
\newcommand{\ablation}{Ablation\xspace}

%%% methods
\newcommand{\ltr}{LTR\xspace}
\newcommand{\ourAlg}{\rforest}
\newcommand{\psvm}{PSVM\xspace}
\newcommand{\lsvm}{LSVM\xspace}
\newcommand{\lreg}{LReg\xspace}
\newcommand{\rforest}{RForest\xspace}

%%% collections
\newcommand{\asr}{ASR\xspace}

%%% metrics
\newcommand{\map}{MAP@5\xspace}
\newcommand{\precFive}{p@5\xspace}
\newcommand{\ndcgFive}{NDCG@5\xspace}
\newcommand{\FOne}{F1\xspace}
\newcommand{\acc}{Acc\xspace}

\newcommand{\scoreFn}[1]{score(#1)}
\newcommand{\rankFn}[2]{rank(#1,#2)}

%%% significance
\newcommand{\statSymbolPSVM}{\star}

%%% table formatting
\newcommand{\negSpace}{\;\;\:}
\newcommand{\statSpace}{\;\;}

%%% ranking function features
\newcommand{\lmirDir}{LMIR.DIR\xspace}
\newcommand{\queryRatio}{QueryTermsRatio\xspace}
\newcommand{\ent}{Entropy\xspace}
\newcommand{\sw}{StopwordsRatio\xspace}
\newcommand{\simToInit}{SimInit\xspace}

%%% features
\newcommand{\metaFt}{Macro\xspace}
\newcommand{\metaFtNoX}{Macro}
\newcommand{\atomicFt}{Micro\xspace}
\newcommand{\atomicFtNoX}{Micro\xspace}

\newcommand{\queryFt}{Query\xspace}
\newcommand{\queryFtNoX}{Query}
\newcommand{\stopwordsFt}{Stopwords\xspace}
\newcommand{\stopwordsFtNoX}{Stopwords}
\newcommand{\noQueryStopsFt}{$\urcorner$Query$\urcorner$Stopwords\xspace}
\newcommand{\noQueryStopsFtNoX}{$\urcorner$Query$\urcorner$Stopwords}
\newcommand{\docNow}{D\xspace}
\newcommand{\docPrev}{PD\xspace}
\newcommand{\winnerPrev}{PW\xspace}
\newcommand{\fOne}{SIM(\docNow,\docPrev)\xspace}
\newcommand{\fTwo}{SIM(\docNow,\winnerPrev)\xspace}
\newcommand{\fThree}{SIM(\docPrev,\winnerPrev)\xspace}
\newcommand{\addOne}{ADD(\winnerPrev)\xspace}
\newcommand{\addTwo}{ADD($\urcorner$\winnerPrev)\xspace}
\newcommand{\rmOne}{RMV(PW)\xspace}
\newcommand{\rmTwo}{RMV($\urcorner$PW)\xspace}

\newcommand{\myparagraph}[1]{\vspace{0.6\baselineskip}\noindent{\textbf{#1}}.~}

\newcommand\platformName{\textit{CSP}\xspace}
\newcommand\platform{\textsc{Simulator}\xspace}
\newcommand\analyzer{\textsc{Analyzer}\xspace}
\newcommand\compare{\textsc{Compare}\xspace}
\newcommand\LlamaEfiveListwiseFour{Llama-E5-Listwise-4-players}
\newcommand\LlamaEfivePairwiseFour{Llama-E5-Pairwise-4-players}
\newcommand\ASRC{Mordo et al.\xspace}
\newcommand\LlamaEfiveListwise{Llama-E5-Listwise}
\newcommand\LlamaEfivePairwise{Llama-E5-Pairwise}
\newcommand\LlamaEfiveListwiseNoCopy{Llama-E5-Listwise-no-copy}
\newcommand\GemmaEfiveListwise{Gemma-E5-Listwise}
\newcommand\LlamaContListwise{Llama-Contriever-Listwise}
\newcommand\LlamaContPairwise{Llama-Contriever-Pairwise}
\newcommand\LlamaOkapiListwise{Llama-Okapi-Listwise}
\newcommand\LlamaOkapiPairwise{Llama-Okapi-Pairwise}
\newcommand\kq{$\kappa_{q}$}
\newcommand\kr{$\kappa_{r}$}
\newcommand\llew{$L \leq W$}
\newcommand\lgw{$L > W$}
\newcommand{\w}{$W$}
\newcommand{\LL}{$L$}
\newcommand{\shared}{Instructional Part}
\newcommand{\contextualized}{Contextualized Part}

\title{CSP: A Simulator For Multi-Agent Ranking Competitions}

% \settopmatter{printacmref=true, printfolios=false}
% \settopmatter{printacmref=fa/lse}
% \fancyhead{}
\author{Tommy Mordo}
\affiliation{%
  \institution{Technion}
  \city{Haifa}
  \country{Israel}
}
\email{tommymordo@technion.ac.il}

\author{Tomer Kordonsky}
\affiliation{%
  \institution{Technion}
  \city{Haifa}
  \country{Israel}
}
\email{tkordonsky@campus.technion.ac.il}

\author{Haya Nachimovsky}
\affiliation{%
  \institution{Technion}
  \city{Haifa}
  \country{Israel}
}
\email{haya.nac@campus.technion.ac.il}

\author{Moshe Tennenholtz}
\affiliation{%
  \institution{Technion}
  \city{Haifa}
  \country{Israel}
}
\email{moshet@technion.ac.il}

\author{Oren Kurland}
\affiliation{%
  \institution{Technion}
  \city{Haifa}
  \country{Israel}
}
\email{kurland@technion.ac.il}

\renewcommand{\shortauthors}{Mordo et al.}

\begin{abstract}
In ranking competitions, document authors compete for the highest rankings by modifying their content in response to past rankings. Previous studies focused on human participants, primarily students, in controlled settings. The rise of generative AI, particularly Large Language Models (LLMs), introduces a new paradigm: using LLMs as document authors. This approach addresses scalability constraints in human-based competitions and reflects the growing role of LLM-generated content on the web—a prime example of ranking competition.
We introduce a highly configurable ranking competition simulator that leverages LLMs as document authors. It includes analytical tools to examine the resulting datasets. We demonstrate its capabilities by generating multiple datasets and conducting an extensive analysis. Our code and datasets are publicly available for research.
\end{abstract}
\maketitle

\section{Introduction}
Ranking incentives can drive corpus dynamics in competitive search
settings \cite{kurland_competitive_2022}: document authors
(publishers) might respond to rankings induced for queries of interest
by modifying their documents. The goal of the modification is to
improve future ranking. A case in point, in Web search, the documents most highly ranked often attract most clicks \cite{Joachims+al:05a}. Hence, for queries of commercial intent, high ranks are of utmost importance.

This practice of modifying documents to improve their future ranking
is often referred to as search engine optimization (SEO)
\cite{Gyongyi+Molina:05a}. Competitive search
\cite{kurland_competitive_2022}, which is our focus in this paper, refers
to white hat SEO: legitimate modifications that do not hurt content
quality and/or the search ecosystem.

There are a few recent studies of competitive search. Ben Basat et
al. \cite{Basat+al:17a} showed using game theoretic analysis that the
probability ranking principle (PRP) \cite{Robertson:77a} is not
optimal in competitive search: it leads to reduced topical diversity
in the long run. The implication is that the static view of a
corpus in most work on ad hoc retrieval falls short given the dynamics
driven by ranking incentives. Raifer et
al. \cite{raifer_information_2017} showed using theoretical and
empirical analysis that a prevalent strategy of publishers which leads to an equilibrium is to mimic
content in the documents most highly ranked in the past. Indeed, previous rankings are the only signal about the undisclosed ranking function.
Goren et
al. \cite{goren_driving_2021} empirically showed that this strategy
leads to a publisher herding effect with potentially unwarranted implications; e.g., reducing the volume of content relevant to a query in the corpus.

Performing empirical studies of competitive search is an extremely
difficult challenge \cite{kurland_competitive_2022}. For example, in
the Web setting, multiple factors can affect the changes of Web pages,
many of which may not be due to ranking incentives. Hence, while there
are observational studies of Web dynamics (e.g.,
\cite{Radinski+al:13a}), analyzing ranking-incentivized publisher
strategies and corpus effects remains an open question. This challenge
drove forward a new type of empirical analysis in recent work on
competitive search: controlled ranking competitions were held between
students
\cite{raifer_information_2017,goren_ranking-incentivized_2020,goren_driving_2021,nachimovsky_ranking-incentivized_2024}. The
students modified documents to have them highly ranked for queries
they were assigned to. They competed over a few rounds, in each of
which they were shown the last ranking induced by the undisclosed
ranking function. The resultant datasets were used for various
analyses
\cite{raifer_information_2017, goren_ranking-incentivized_2020, goren_driving_2021,Wu+al:23a,nachimovsky_ranking-incentivized_2024}.

Ranking competitions with humans acting as publishers are valuable for
offline analysis of the {\em specific} competitions that took place. However,
studying new retrieval methods and/or publishers' document
modification strategies, which significantly affect the competition, is
practically impossible: each new design choice calls for re-running the
competition. Kurland and Tennenholtz \cite{kurland_competitive_2022} mentioned this challenge as a potential barrier to empirical study of competitive search and suggested to run ranking competitions using automated agents. Their call of arms, together with the increasing proliferation over the Web of generated AI content, is the motivation for the work we report here.

We present a platform dubbed \platformName: a multi-agent platform
that uses large language models (LLMs) as publishers (agents) in
ranking competitions. The platform allows the execution of large-scale
and highly varied ranking competitions where various factors can
be controlled: the query for which the competition is held, the
ranking function, the LLM and its prompt. The platform also provides a
competition Analyzer module that analyzes a single competition using
various measures. In addition, the platform includes a Compare module that
allows to compare different competitions; e.g., competitions run with different LLMs or different rankers. The \platformName platform will be made publicly
available upon acceptance of this paper, and is currently available
for reviewing purposes at
{\url{https://github.com/csp-platform/Simulator}.}

To demonstrate the merits of \platformName, we used it to run many
ranking competitions for sets of queries where we varied the LLM and
its prompt and the ranking function. We present analysis of the
resultant competition datasets using the Analyzer and Compare modules. For example, we
compare the corpus dynamics with that reported in past human-based
ranking competitions and reveal that LLM-based agents reduce content diversity in the corpus to a larger extent than humans. Furthermore, the analysis of the competitions shows that the ranking function has less effect on the
dynamics than the choice of the LLM which serves as a publisher.
% \begin{figure}[t]
%     \centering
%     \includegraphics[width=0.8\linewidth]{figs/platform/ecosystem.jpg}
%     \caption{Components of the \textsc{\platformName} framework: (1) \textsc{Platform} simulates a ranking competition; (2) \textsc{Analyzer} analyzes a single competition; and (3) \textsc{Compare} performs comparisons across multiple competitions.}
%     \label{fig_lemss_ecosystem}
% \end{figure}
% motivation:
% \begin{itemize}
%     \item help IR designers to consider different aspects of a IR system
%     \item how llms compete in IR ecosystems
% \item simulate how LLMs behave in competitive search settings.
% \end{itemize}

\section{Related Work}
Previous studies addressed theoretical and empirical results. Kurland and Tennenholtz \cite{kurland_competitive_2022} highlight several perspectives in the competitive search ecosystem. From \textit{the ranker perspective}, the focus is on designing mechanisms that can enhance social welfare (i.e., mitigate herding of publishers and promote fairness among publishers \cite{kurland_competitive_2022}). \textit{The publisher perspective} includes benefits to document authors who strategically manipulate their documents to improve the future rankings. \textit{The user perspective} includes benefit of the user, who interacts with the search engine to satisfy an information need expressed using a query. To empirically explore these perspectives, ranking competitions in various settings have been conducted and analyzed.

\myparagraph{Ranking competitions} Game theory \cite{aumann1995repeated, fudenberg_game_1991} provides a foundational framework for modeling repeated ranking competitions where document authors do not cooperate.

Several studies have used controlled ranking competitions as a tool to study dynamics in competitive settings. Raifer et al. \cite{raifer_information_2017} conducted the earliest controlled ranking competitions among students, revealing a key strategy employed by publishers: mimicking the documents most highly ranked in previous rounds. Goren et al. \cite{goren_driving_2021} demonstrated how search engines could drive predefined and targeted content effects, leading to the herding phenomenon \cite{banerjee_simple_1992}. Nachimovsky et al. \cite{nachimovsky_ranking-incentivized_2024} extended this line of research by conducting ranking competitions where authors competed to improve their ranks with respect to multiple queries.

In addition to these studies, another line of research focuses on developing algorithmic adversarial attacks designed to manipulate rankings and promote specific documents. Examples include Castillo et al.'s work on adversarial strategies \cite{castilo_adverserial}, Raval and Schwing's exploration of one-shot adversarial attacks \cite{raval_one_2020} and Mordo et al.'s study on strategic document manipulation in competitive search setting with diversity-based ranking \cite{mordo_search_2025}.

\myparagraph{LLMs in competitive search} The use of large language models (LLMs) in competitive search setups has several aspects. First, a growing number of document authors leverage LLMs to modify and even generate content for search engine optimization (SEO) purposes \cite{Memon2024-re, Aggarwal2023-aw, nachimovsky_ranking-incentivized_2024, wu_survey_2024}. LLMs can generate high-quality content \cite{nachimovsky_ranking-incentivized_2024}, sometimes indistinguishable from content generated by humans \cite{wu_survey_2024}, and aimed at improving rankings. Second, various tasks require data annotation are not scalable, and can benefit from LLM assistance \cite{tan_large_2024, faggioli_perspectives_2023}. Organizing ranking competitions among human participants, similar to conducting annotation tasks, has traditionally posed significant challenges of scalability. LLMs provide new opportunities for studying ranking competitions by simulating document authors, enabling researchers to simulate competitive search scenarios efficiently.

% LLMs provide a compelling alternative to human participants, enabling researchers to simulate competitive search scenarios efficiently. For instance, LLMs can emulate agents engaging in "white hat" optimization tactics or facilitate the design of optimal ranking functions to promote social welfare.
Third, the emergence of LLMs has introduced novel dynamics in human-LLM interactions, particularly within competitive search settings \cite{nachimovsky_ranking-incentivized_2024, bardas_prompt-based_2025}. For example, researchers can study how document authors react to LLM-generated content \cite{nachimovsky_ranking-incentivized_2024}, or explore how LLMs perform when competing against human players in ranking games \cite{bardas_prompt-based_2025}. Finally, LLMs themselves are the basis for the potential shift from search engines to (conversational) question answering (QA) systems (e.g., \cite{perplexity}). Additionally, these systems can be used for zero-shot or few-shot ranking tasks \cite{Hou2024-tm, Qin2023-hg, bardas_prompt-based_2025}.

% Finally, LLMs themselves are increasingly being employed as ranking functions. For instance: Encoders such as E5 \cite{wang_text_2024} and Contriever \cite{izacard_unsupervised_2022}; chatbot-based systems such as ChatGPT used for 

\myparagraph{Agents in decision-making tasks} Recent research has increasingly highlighted the capabilities of large language models (LLMs) in decision-making tasks, demonstrating their potential to function as autonomous agents in complex economic environments that often require advanced strategic reasoning \cite{horton_large_2023, wang_text_2024, zhu_capturing_2024, li_stride_2024, shapira_glee_2024, noauthor_letta_nodate}. In other work, LLMs were used to simulate users in information retrieval and recommendation systems \cite{breuer_report_2024, zhang_generative_2024, wang_user_2024}, and we use them to simulate authors.

\myparagraph {LLM-based multi-agent simulation} Various tools and frameworks have been developed to create agents powered by LLMs\footnote{For example, \url{https://www.letta.com} and \url{https://github.com/Thytu/Agentarium/tree/main}} \cite{feng_agile_2024, qiao_taskweaver_2024, liu_agentlite_2024}. However, these tools are not easily customizable to simulate ranking competitions. They lack specific modules designed to facilitate the analysis and exploration of unique aspects of ranking competitions, such as dynamic interactions between agents and ranking strategies.

\section{{\platformName} framework}
\subsection{Competitive search background}\label{sec_background}
The \platformName \platform supports the simulation of ranking competitions. We now turn to describe the structure of ranking competitions, and more generally, competitive search \cite{kurland_competitive_2022}.

A ranking game between publishers (document authors) is driven by their ranking incentives. That is, we assume that some publishers are motivated to have their documents highly ranked for a query. In response to a ranking induced for the query, the publishers (players) might modify their documents so as to improve their future ranking. In the general setup \cite{raifer_information_2017}, every player begins with an \textit{initial document} pre-determined by the system designer. During each round of the game, every player: (1) modifies her document, and (2) receives information about the ranking induced for the query; the player can observe the content of other documents in the ranking. Players may then modify their documents in subsequent rounds with the goal of promoting their rankings with respect to the query of the game. We refer to a collection of games, each associated with a different query, as a {\em competition}.

% \ref{fig_scheme_game}.

% \begin{figure}[htbp]
%     \centering
%     \includegraphics[width=0.8\linewidth]{figs/platform/shceme_game_2.jpg}
%     \caption{Scheme of a ranking game.}
%     \label{fig_scheme_game}
% \end{figure}

% Each competition simulated on our platform comprised a batch of games. Each game is associated with a query and begins with an initial document containing information relevant to that query. A set of agents is assigned to participate in the game. During each round, players receive feedback on the documents published by other players and modify their candidate documents accordingly to achieve the highest ranking.

\subsection{{\platformName} {\platform}}
The core component of {\platformName} is the {\platformName} {\platform}, a highly configurable simulator designed to simulate ranking competitions.
% This simulator enables the study of repeated ranking games, where document authors (players) compete for a query, and iteratively modify their document to promote their rankings for an undisclosed ranking function.
It allows fine-grained control over key parameters, including the ranking function, the query for which every game is performed, initial documents, and the type of the players: a specific type of a player is referred to as an {\em agent}. Specifically, \platformName{} is highly supports players based on LLMs. Additionally, it supports the configuration of the number of simulated games, the number of players per game and the number of rounds per game, providing flexibility for diverse experimental setups. The simulation output includes a dataset containing all documents generated by players across rounds, along with their rankings, forming the basis for competition analysis.

% Formally, our framework focuses on three primary aspects: \textit{game properties}, \textit{agent properties}, and \textit{ranking function properties}.

% \textit{Game properties} are the parameters of games, such as the number of rounds in every repeated game, the number of players in each game, and other relevant settings. (Further details are provided in Section \ref{platform_game_properties}.)

% \textit{Agent properties} refer to as the characteristics of the agents, including the agent type (e.g., LLM-based agent or human agent), the structure of the message she received by the system designer, and additional behavioral traits; more information on \textit{Agent properties} can be found in Section \ref{platform_agent_properties}).

% \textit{Ranking function} properties include the type of the ranking function that induces the ranking of players' documents in each round of the games (more details are provided in Section \ref{platform_ranking_properties}).

The simulation parameters can be configured via a \textit{json} file or customized directly within the {\platformName} {\platform} code-base, offering significant flexibility for tailoring the platform to various experimental scenarios. The code-base is publicly available\footnote{\url{https://github.com/csp-platform/Simulator}.}.

% In section \ref{section_data}, we demonstrate the platform's capabilities by generating datasets from various types of competitions involving LLMs. Additionally, in section \ref{section_analysis}, we present an array of methods for analyzing and comparing competitions to gain insights into competition's dynamics. We used those methods to compare different types of competitions.

\subsubsection{Competition properties}\label{platform_game_properties}

A ranking competition is structured as a collection of games. Each game is composed of a set of players competing over a few rounds for a predefined query. A competition is highly configurable, allowing customization of parameters such as the assigned query for every game, and the corresponding initial document, the number of rounds and more.
% The platform supports two operational modes: \textit{round-by-round} and \textit{game-by-game}. In \textit{round-by-round} mode, each round of the competition involves all games being played simultaneously, with players modifying their documents once per round. This mode can accommodate online scenarios, such as competitions involving human participants against other humans or against LLMs. In contrast, the \textit{game-by-game} mode involves completing all rounds of a single game before proceeding to the next. This approach offers advantages in optimizing memory allocation efficiency and reducing runtime.
% Our \platformName{} \platform{} has the option to simulate a competition from a predefined state (e.g., after several rounds of each game). This functionality help exploring different scenarios that start from different entry point. Additionaly, it facilitates splitting lengthy simulations into smaller batches for computational and experimental efficiency. 
Additional, our \platformName{} \platform{} offers the capability to simulate a competition from a predefined state (e.g., after several rounds of each game). This functionality aids in exploring various scenarios that begin from different starting points. Moreover, it allows for breaking lengthy simulations into smaller batches.

% Additionally, it enables researchers to explore competitions under varying initial conditions, offering insights into how different starting points influence the outcomes and dynamics of the competition.

% This approach offers advantages in terms of efficiency of memory allocation and reduced runtime.

\subsubsection{Player properties}\label{platform_agent_properties}
We assume that the same set of players is assigned to each game within a competition. Consequently, the only difference between games in a competition is the query being played and the corresponding initial document.
% However, our simulator is versatile and can also enable competitions involving predefined documents or mixed scenarios with LLMs and predefined documents.
Players are exposed to documents from previous rounds along with their corresponding rankings. Note that the rankings are the only signals available to players regarding the undisclosed ranking function. Accordingly, players modify their documents so as to be ranked higher in subsequent rounds. Different players may interpret and utilize these signals in various ways to guide their document modifications. In this paper, we focus on scenarios where different LLMs compete against each other. 

\subsubsection{Ranking function properties} \label{platform_ranking_properties}
The ranking function determines the rankings of players' documents in every round of a game. The {\platformName} {\platform} is designed to be highly flexible, allowing the integration of any type of ranking function such as feature-based or neural-based ranking functions.
Additionally, the {\platformName} {\platform} supports implementing penalty mechanisms to enhance the dynamics of players in the competition. Inspired by ranking competitions among human participants \cite{mordo_search_2025, raifer_information_2017, nachimovsky_ranking-incentivized_2024}, these mechanisms can penalize players—by demoting their rankings—for copying documents from others or for refrain from modifying their documents across multiple rounds.

\subsection{{\platformName} {\analyzer} and {\platformName} {\compare}}
We introduce two additional components in the {\platformName} framework designed to facilitate the analysis of competitions.

The {\platformName} {\analyzer} is a module designed to enable in-depth analysis of an individual competition. The analysis is focused on different measures related to different aspects of the competition dynamics. For example, we measure how the diversity of documents of ranked lists evolves over rounds \cite{mordo_search_2025}.
% The {\platformName} {\analyzer} framework is implemented in a \textit{Python} module that processes competition datasets 
% through the following steps: (1) integrating the dataset of a ranking competition into a background index\footnote{We used Clueweb09 in our demonstration. The integration is needed for collection statistics.} implemented using Pyserini \cite{Lin_etal_SIGIR2021_Pyserini}; (2) extracting document features, such as Okapi BM25 scores; (3) generating document representations, such as TF.IDF and SBERT \cite{reimers_sentence-bert_2019}; and (4) performing analyses on the documents using different document representations and measures. Further details on the measures are formally described in Section \ref{label_subsection_evaluation_measures}.
For cases where comparisons between datasets generated under different competitive scenarios are of interest, we developed {\platformName} {\compare}, an interactive tool (dashboard) for comparing competitions with different configurations.

The functionalities of {\platformName} {\analyzer} and {\platformName} {\compare} are demonstrated using several research questions. (See Section \ref{label_section_RQs}.) The code-bases are available on GitHub\footnote{{\platformName} {\analyzer}: \url{https://github.com/csp-platform/Analyzer} and {\platformName} {\compare}: \url{https://github.com/csp-platform/Compare}}. Note that {\platformName} {\analyzer} and {\platformName} {\compare} were also designed to analyze and compare ranking competitions, even if they were not generated by the {\platformName} {\platform}; e.g., competitions conducted between human players \cite{raifer_information_2017, goren_driving_2021, nachimovsky_ranking-incentivized_2024}.
\section{Creating Datasets Using \platformName}\label{section_data}

We created 22 datasets using the {\platformName} {\platform}, each produced through a simulation representing a distinct configuration of a ranking competition among LLMs.
We analyzed the datasets using the {\platformName} {\analyzer} and the {\platformName} {\compare} modules to demonstrate an analysis of a single competition and comparison of multiple competitions. (The analysis is presented in Section \ref{section_analysis}.)
Additional objective of collecting data from LLM-based competitions is to compare the properties of ranking competitions conducted among LLMs with those involving human players. To this end, we utilized the dataset of a ranking competition conducted between students, organized by Mordo et al. \cite{mordo_search_2025}. The competition included 15 games, each for a different query selected from TREC09-TREC12. Every game included four students and lasted for seven rounds. The ranking function was the cosine between a query and document embedding induced using E5 \cite{wang_text_2024}. The competition was approved by ethics committees \cite{mordo_search_2025}.

% A penalty mechanism was implemented to encourage players to actively modify their documents and to discourage copying of other documents.

% The dataset includes: (1) the documents of the competition; (2) the ranking of every document; (3) the relevance judgments; and (4) the quality judgments.

\subsection{The resultant datasets of LLM-based competitions}
The parametrization of the {\platformName} {\platform} enables a diverse and flexible range of options for simulating ranking competitions and generating their corresponding datasets. Our goal in this paper is to demonstrate {\platformName}' capabilities and not to find the best performing agents. We adopt a specific class of prompt-based agents developed by Bardas et al. \cite{bardas_prompt-based_2025}. Note that other types of LLMs (for instance, fine-tuned LLMs) can be integrated in the {\platformName} {\platform}. The proposed agents are LLMs that operate based on instructions provided through prompts. Each agent is defined by a specific LLM model and a prompt given each round, that guides its behavior in the ranking game. Each prompt has two parts: a general shared part which describes the task (denoted \textit{\shared}) and a context-specific part, specific to a prompt, which provides information about past rankings (denoted \textit{\contextualized}).
 % Two components: (1) the foundational model of the LLM, along with its specific parameters (e.g., Llama3.1 \cite{dubey_llama_2024} with temperature=0.5, $top_p = 0.9$, etc); (2) The prompts used to instruct the LLM.
 % Two types of prompts were discussed in Bardas et al.\cite{bardas_prompt-based_2025}: \textit{System prompt} and \textit{User prompt}. 

The \shared{} includes the general background for the game, which outlines the game's rules and restrictions, the player's current document, and the assigned query. Additionally, the \shared{} instructs players to maintain similarity to their original document as a guideline to ensure faithfulness to the original content.
% For more details see Figure \ref{prompt_system} in Appendix \ref{appendix_prompts}.
% In competitions conducted among students \cite{raifer_information_2017, nachimovsky_ranking-incentivized_2024, mordo_search_2025}, a behavior of mimicking the winner was observed, where authors of lower-ranked documents copied content from top-ranked documents. This behavior negatively impacted the dynamics of the game, leading to a reduction in topical diversity \cite{goren_driving_2021}.
% competition organizers implemented penalty mechanisms, such as reducing the bonus points students could earn for participating. Since our setting involves LLMs as players,
To mitigate the potential behavior of \textit{mimicking-the-winner} (mimicking content from the top-ranked documents) as observed in competitions between students \cite{raifer_information_2017}, we introduced a modified \shared{}. It is building upon the version presented by Bardas et al. \cite{bardas_prompt-based_2025}, with an additional instruction explicitly designed to discourage general copying of content during the game. This mitigation is crucial, as the long-term effect of \textit{mimicking-the-winner} can lead to a herding effect, ultimately reducing topical diversity in the corpus and limiting the competitive dynamics of the game \cite{goren_driving_2021}.
% (see Figure \ref{prompt_system_no_copy} in Appendix \ref{appendix_prompts}).

The \contextualized{} of the prompt includes information about past rankings, to serve as a signal about the ranking function. (Recall that the ranking function is not disclosed.) It guides the LLM's actions in subsequent rounds to try to achieve the highest ranking. Bardas et al. \cite{bardas_prompt-based_2025} conducted a series of experiments aimed to find the optimal parametrization of the \contextualized s which maximizes ranking promotion. We use the two best performing \contextualized s reported by Bardas et al. \cite{bardas_prompt-based_2025}. These prompts led to the highest ranking promotion for the modified documents: Pairwise prompt and Listwise prompt. The Pairwise prompt contains pairs of randomly selected documents and their rank preference by the ranking function, from the last three rounds. The Listwise prompt contains ranked lists from the two last rounds.

We demonstrate the platform's capabilities using a competition among LLMs.
% A summary of the parameters defining the configuration of each competition and the resultant datasets appear in Appendix \ref{appendix_dataset_parameterts} Tables \ref{table_data_param} and \ref{table_datasets}.
Each competition consisted of 30 games, with each game held for a query with a commercial intent, selected from the TREC09–TREC12 datasets\footnote{We used the following queries: 9,17,21,29,34,45,48,55,59,61,64,69,74,75,78,83,96,98,124,144,\\ 161,164,166,167,170,180,182,193,194,195 from \url{https://trec.nist.gov/data/webmain.html}.}. The initial documents, those provided to the players at the beginning of \ASRC's dataset \cite{mordo_search_2025}.

Each game was played for 30 rounds. We used different lightweight instruct-tuned (<10B parameters) language models as agents: Llama 3.1 \cite{dubey_llama_2024}, Gemma2 \cite{gemma_team_gemma_nodate}, Qwen2.5 \cite{qwen_qwen25_2024} and Ministral\footnote{meta-llama/Meta-Llama-3.1-8B-Instruct, google/gemma-2-9b-it, Qwen/Qwen2.5-7B-Instruct and mlx-community/Ministral-8B-Instruct-2410-bf16 from Hugging face repository.} \cite{jiang_mistral_2023}.
For demonstration purposes, we use Llama3.1 and Gemma2 in the analysis presented in Section \ref{section_analysis}, and prompt them with the Listwise and Pairwise prompts. Due to computational limitations, each competition utilized only a single language model and \contextualized{}, with the same commonly used hyper-parameters \cite{li_examining_2024}: a $top_p$\footnote{Controls the probability mass from which the model samples its next token, ensuring that only the most probable tokens, comprising 90\% of the cumulative distribution, are considered.} value of 0.9 and a temperature\footnote{Controls the randomness of language model output.} of 0.5, enabling LLMs to generate diverse content while prioritizing the most probable tokens. To introduce variation among competing players, each player was assigned a persona \cite{shapira_can_2024,samuel_personagym_2024}. This approach aimed to induce diverse behavioral patterns among the players in the same game. The personas were generated as follows: we selected the Educational environment from Samuel et al. \cite{samuel_personagym_2024} to align with the topic interests of students, given that previous ranking competitions were conducted between student. Then, we used GPT-4o to generate five personas using the prompt provided by Samuel et al. \cite{samuel_personagym_2024} with an additional general description of ranking competitions. The resulting personas included a BSc student, a professional writer, a professional editor, an English teacher, and a Data Science professor.

Four or five players competed against each other in each game. Three distinct unsupervised ranking functions were applied. Two dense retrieval methods and one sparse method: E5 \cite{wang_text_2024}, Contriever\footnote{intfloat/e5-large-unsupervised and facebook/contriever from Hugging face repository.} \cite{izacard_unsupervised_2022} and Okapi BM25 \cite{maron_relevance_1960}. To compute the score in dense retrieval methods for a given document-query pair, we first generated the embedding vectors for both the document and the query using the embedder model, then computed the cosine similarity between these vectors. For extracting IDF (inverse document frequency) based features in Okapi, we used the English Wikipedia dataset with 59k pages, from a 2020 dump (Krovetz stemmed) \cite{Frej2020Wikir,Frej2020MlWikir}.

The LLMs sometimes generate text that exceeds the word limit allowed in the \shared{} or includes unnecessary headers and prefixes\footnote{For instance: "The modified document is:" or "Here is the document text:".}. To prevent exceeding token limits, we truncated the generated documents to the first 256 tokens\footnote{To align with competitions conducted between students \cite{raifer_information_2017, nachimovsky_ranking-incentivized_2024, mordo_search_2025}.}. To address the presence of unwarranted tokens, we incorporated a post-edit step in which players utilize their own LLM with a targeted cleaning prompt. This prompt directs the LLM to remove any headers or prefixes from the generated document.

% , ensuring that only the main part of the document is extracted prior to submission to the ranker.

% The prompt used for this cleaning process is detailed in Figure \ref{prompt_clean} in Appendix \ref{appendix_prompts}.

The competitions with five players per game resulted in 4500 documents, and those with four players per game yielded 3600 documents.
% Recall, the goal of this research is not to identify the optimal agents that achieve the highest ranking, as explored in Bardas et al. \cite{bardas_prompt-based_2025}. Instead, we focus on demonstrating the platform's functionality, leaving the exploration of optimization objectives for future work.
\section{Analysis of Competition Datasets}\label{section_analysis}
In this section, we present an extensive analysis of seven datasets (out of the 22 created datasets) generated using the {\platformName} {\platform}, selected due to space constraints. All the datasets and their corresponding analyses, conducted using the {\platformName} framework, are available in our GitHub repository\footnote{\url{https://github.com/csp-platform/Datasets}.}. In Section \ref{label_section_RQs} we present three research questions used to compare the seven datasets. Then, in Section \ref{label_subsection_evaluation_measures} we discuss the measures used to analyze the datasets and to compare them. The analysis addressing the three research questions is detailed in Sections \ref{section_rq1}, \ref{section_rq2}, and \ref{section_rq3}. In this section, we define a "winner" as the author of the top-ranked document in a given round, or the document itself. Conversely, "losers" refer to authors (or documents thereof) whose documents did not achieve the highest rank in that round.

\subsection{Research questions}\label{label_section_RQs}
We address three main research questions:
\begin{itemize}
    \item (RQ1) What are the differences between competitions with LLMs as players and competitions with humans as players in terms of player's strategies, resulting documents, and more? To address this question, we utilized \ASRC's dataset \cite{mordo_search_2025} described in Section \ref{section_data}.
    \item (RQ2) How do the characteristics of LLM-based agents (specifically, the language model and prompt type) affect the dynamic of the competition? For this analysis, we fixed the ranking function to the (unsupervised) E5 \cite{wang_text_2024}.
    \item (RQ3) How does the choice of ranking function affects the competition dynamics? For this comparison, we fixed the LLM to Llama3.1 \cite{dubey_llama_2024}.
\end{itemize}
A summary of the datasets used for addressing these research questions is presented in Table \ref{label_table_RQs}. Additionally, for every research question, we focused on a subset of measures, prioritizing those that provided the most insightful findings. We emphasize that overall our findings are consistent among all measures.

% All datasets and the full comparative results are publicly available in our GitHub repository.

\myparagraph{Quality and relevance judgements}
We annotated the datasets used for addressing the three research questions. (See Table \ref{label_table_RQs}.) Each document was judged for binary relevance to a query by three crowd workers (English speakers) on the Connect platform via CloudResearch \cite{noauthor_introducing_2024}. Three workers annotated the quality of each document with the categories: valid, keyword-stuffed\footnote{Adding query terms to documents in an excessive manner.}, and spam. We used the same instructions to annotators as presented in past work \cite{nachimovsky_ranking-incentivized_2024, mordo_search_2025}. For each document, the final quality grade was defined as the number of annotators who judged the document as valid. Accordingly, the final relevance grade was the number of annotators who marked the document as relevant. Due to budget limitations, annotations were not performed on datasets where the Contriever \cite{izacard_unsupervised_2022} ranking function was used. Additionally, we annotated only the top-ranked document from every two rounds. The inter-annotator agreement rates (free-marginal multi-rater Kappa \cite{fleiss_measuring_1971}) ranged between 45.2\% and 66.8\% for quality judgments. For relevance judgments, the agreement rates ranged from 46.5\% to 66.0\%. Since in \ASRC \cite{mordo_search_2025}, every document was annotated by five workers, we calibrated the relevance and quality grades to 3 by multiplying the relevance and quality grades of \ASRC{} \cite{mordo_search_2025} by $\frac{3}{5}$. In addition, we selected from \ASRC \cite{mordo_search_2025} only the top-ranked documents to align with the comparison with the LLM-based competitions here.

\begin{table}[t]
\caption{Research questions (RQs) and the corresponding competition datasets used to address them. Each competition includes the following components: either Llama or Gemma-LLM; a ranking function (E5, Contriever, or Okapi); Listwise or Pairwise \contextualized{} of the prompts; and four or five players per game (unless specified otherwise, the competition includes five players). One of the competitions in RQ2 includes a modified \shared{} that instructs the LLM not to copy other players' documents (denoted as 'no-copy'). We annotated all the listed datasets, excluding datasets with a Contriever ranking function. \kq, \kr{} are the inter-annotator agreement rates (free-marginal multi-rater Kappa) of the quality and relevance judgements, respectively.}
\label{label_table_RQs}
\centering
\begin{tabular}{|c|l|c|}
\hline
\textbf{RQ} & \textbf{Dataset} & \textbf{Annotation (\kq, \kr)} \\ \hline
{\textbf{RQ1}} 
& \LlamaEfiveListwiseFour & (58.5\%, 65.3\%) \\ \cline{2-3}
& \LlamaEfivePairwiseFour & (51.4\%, 53.8\%) \\ \cline{2-3}
& \ASRC \cite{mordo_search_2025} & (32.0\%, 62.3\%) \\ \hline

{\textbf{RQ2}}
& \LlamaEfiveListwise & (66.8\%, 66.0\%) \\ \cline{2-3}
& \LlamaEfivePairwise & (47.3\%, 48.4\%) \\ \cline{2-3}
& \LlamaEfiveListwiseNoCopy & (53.8\%, 57.3\%) \\ \cline{2-3}
& \GemmaEfiveListwise & (71.8\%, 61.2\%) \\ \hline
{\textbf{RQ3}} 
& \LlamaEfiveListwise & (66.8\%, 66.0\%) \\ \cline{2-3}
& \LlamaEfivePairwise & (47.3\%, 48.4\%) \\ \cline{2-3}
& \LlamaContListwise & - \\ \cline{2-3}
& \LlamaContPairwise & - \\ \cline{2-3}
& \LlamaOkapiListwise & (60.9\%, 60.3\%) \\ \cline{2-3}
& \LlamaOkapiPairwise & (45.2\%, 46.4\%) \\ \hline
\end{tabular}
\end{table}

\subsection{Measures}\label{label_subsection_evaluation_measures}
We propose a broad set of measures applied to the competitions datasets. These measures were designed to facilitate the analysis of competitions and to gain multifaceted insights into the dynamics and outcomes of competitive search scenarios. To quantify similarity between documents we employ the TF.IDF\footnote{The results obtained using alternative text representations, including E5 \cite{wang_text_2024} with cosine similarity and SBERT \cite{reimers_sentence-bert_2019} with cosine similarity, were consistent with those attained for TF-IDF. Thus, they are omitted for brevity.} text representation with cosine similarity. Note that additional text representations can be easily integrated into \platformName and may, in some cases, provide different insights \cite{mordo_search_2025}. The measures are averaged over queries (games) unless stated otherwise.

We categorized the measures to five classes of competition properties:

% we employ various text representations, including (unsupervised) E5\footnote{intfloat/e5-large-unsupervised.} \cite{wang_text_2024} representation with cosine similarity; SBERT \footnote{all-MiniLM-L6-v2.} \cite{reimers_sentence-bert_2019} representation with cosine similarity; and JACCARD similarity

% (1) mimicking the winner; in past work the strategy of mimicking the top-ranked documents ("winners") was observed, and formally proofed as the optimal strategy of players in ranking games  \cite{raifer_information_2017}. (2) Diversity of the ranked list of documents; inspired by the work of Mordo et al. \cite{tommy}(3) Convergence of the competition; i.e., analyze the modifications of documents in consecutive rounds.(4) Relevance and Quality; and (5) Winner (author of top-ranked document in a ranked list) Properties.
% The evaluation methods measures corresponding to each class of properties are fully implemented within the {\platformName} {\analyzer}. 
\begin{itemize}
    \item \textit{\textbf{Mimicking-the-winner}} (mimicking content from top-ranked documents) has been identified, both theoretically and empirically, as a strategy employed by players who loose in a round to improve their chance of winning in subsequent rounds \cite{raifer_information_2017}. Following prior work, we analyzed the document modification strategies by analyzing the changes in the documents' feature values\footnote{The features are selected from \url{http://www.research.microsoft.com/en-us/projects/mslr} or from past work on ranking competitions \cite{raifer_information_2017, nachimovsky_ranking-incentivized_2024, mordo_search_2025}.} over rounds \cite{qin_introducing_2013,nachimovsky_ranking-incentivized_2024, raifer_information_2017, mordo_search_2025}. The background dataset used to compute IDF based features was Clueweb09\footnote{\url{https://lemurproject.org/clueweb09/}.}.
    We analyzed changes in feature values of winner documents between consecutive rounds; $W_i$ and $W_{i+1}$ are the
winner documents in rounds $i$ and $i+1$, respectively. We focus on cases where $W_i$ and $W_{i+1}$ are produced by different
players as it was observed in past work
\cite{raifer_information_2017, mordo_search_2025} that players who win a round are
unlikely to substantially modify their document for the subsequent round.

% on changes in documents that lost for at least three consecutive rounds before they reach the first position. We compare the average feature values of winners in each corresponding round (\w) with those of the loser  (\LL). Inspired by Raifer et al. \cite{raifer_information_2017}, we split the losers into two groups according to whose feature values were lower than or equal to the winner (\llew); or greater than the winner (\lgw), three rounds prior to their win.
Additional approach to studying the phenomenon of \textit{mimicking-the-winner} is to analyze the similarity of winners' documents between consecutive rounds \cite{mordo_search_2025}. As in the feature values analysis, we focus on cases where $W_i$ and $W_{i+1}$ are produced by different players. Note that high similarity values indicate that a player has made modifications that make her document similar to the previously winning document. We also analyze the similarity between the top two ranked documents as an additional indicator of the herding effect\footnote{The tendency of players' documents to converge towards similar content, leading to a reduction in topical diversity in the ranked list.} \cite{goren_driving_2021}. If the second-ranked player is \textit{mimicking-the-winner}, we expect an increased similarity between the top two ranked documents.
\item \textit{\textbf{Diversity}} of documents. We measure diversity at two levels: the ranked-list-level and the player level. Diversity at the ranked-list-level is computed by the minimum of inter-document similarity (averaged over queries) across rounds \cite{mordo_search_2025}. Diversity at the player level is computed by the similarity of documents produced by the same player between consecutive rounds. It is averaged over players and queries across rounds.
\item \textit{\textbf{Convergence}} of a competition. Informally, this class includes measures that assess whether documents continue to be modified as the game progresses. As in the \textit{Diversity} class of measures, we measure convergence at the ranked-list-level and at the player level. As for the ranked-list-level, we compute the minimum of the inter-document similarity in a ranked list, enabling us to observe how inter-document similarities in a ranked list evolve over rounds. We also examined the number of unique documents over rounds to detect potential convergence to unique documents. At the player level, convergence was estimated by measuring the similarity between documents of the same player between consecutive rounds, as computed in the \textit{Diversity} class of measures. The resulting player level analysis will indicate how players modify their documents in comparison to their documents from the previous round.

\item \textit{\textbf{Quality and relevance}} of documents in a competition. We analyzed the quality and relevance grades of top-ranked documents in each competition across rounds. The annotation process and the methodology for computing the quality and relevance grades are detailed in Section \ref{label_section_RQs}.

\item \textit{\textbf{Top-ranked players' statistics}}. We analyzed two statistics of the top-ranked players in a competition. The first is the proportion of the wins, computed by the ratio of the number of wins by the best-performing player in a game to the average number of expected wins\footnote{The average number of expected wins in a game is computed as the total number of rounds divided by the number of players in a game.} assuming every player has a random chance to win a round of a game (averaged over games). For instance, in our competitions with five players and 30 rounds, one extreme case occurs when each player wins randomly, resulting in a ratio of 1. In the opposite extreme, if a single player wins all 30 rounds in all games, the ratio would be $\frac{30}{6} = 5$. The second statistic is about winning players. We identify the overall best-performing player across all games in a competition (i.e., the player with the highest number of wins, aggregated over rounds and games).
% We analyzed the best agents only on competitions in which LLMs were the agents.
\end{itemize}

\subsection{Results (RQ1): human vs. LLM-based agents}\label{section_rq1}

We now turn to the comparison between competitions with LLM-based agents and those conducted with human players. For the LLM-based competitions, we used the same ranking function and the same number of players per game as used in {\ASRC} \cite{mordo_search_2025}.
% However, no penalty mechanism was applied to maintain alignment with the prompts and setup presented in Bardas et al. \cite{bardas_prompt-based_2025}.
We employed Llama-based models, utilizing both Pairwise and Listwise prompts, along with the \shared{} as described in Bardas et al. \cite{bardas_prompt-based_2025}. Each player was assigned a distinct persona as described in Section \ref{section_data} excluding the persona of a Data Science professor, which was randomly selected to be omitted.

We now turn to study the \textit{mimicking-the-winner} phenomenon. Specifically, we
analyze changes in feature values of winner documents as described in Section \ref{label_subsection_evaluation_measures} \cite{raifer_information_2017, nachimovsky_ranking-incentivized_2024, mordo_search_2025}. The features are divided into two main categories: query-independent and query-dependent features. The query-independent features are (i) \textbf{Length}: document length, (ii) \textbf{StopwordRatio}: the ratio of stopwords to non stopwords in the document; the INQUERY stopword list was used \cite{Allan+al:00a}, and (iii) \textbf{Entropy}: the entropy of the unsmoothed unigram maximum likelihood estimate induced from the document. The query-dependent features are (iv) \textbf{LM.DIR}: the query likelihood score of a document where document language models are Dirichlet smoothed with smoothing parameter set to 1000 \cite{zhai_study_2001}, (v) \textbf{TF}: the sum of query term frequencies in the document, and (vi) \textbf{BM25}: the Okapi similarity between the query and the document. 

The results are presented in Figure \ref{fig_rq1_feature_strategies}. A general decreasing trend in the feature values was observed across rounds in almost all cases, except for Entropy in the human competition and Length in the LLM-based competitions. Notably, the query-dependent feature values exhibited greater differences and a steeper decrease in the human competition compared to the LLM-based competitions. This observation suggests that, similar to human participants \cite{raifer_information_2017}, LLM-based agents may adopt a \textit{mimicking-the-winner} strategy to align with winning features.

\begin{figure}[ht!]
    \centering
    \includegraphics[scale=0.35]{figs/RQ1/graphs_comparison.png}
    \begin{minipage}[t]{0.48\textwidth}
        \centering
        \begin{minipage}[t]{0.32\textwidth}
            \includegraphics[width=\linewidth]{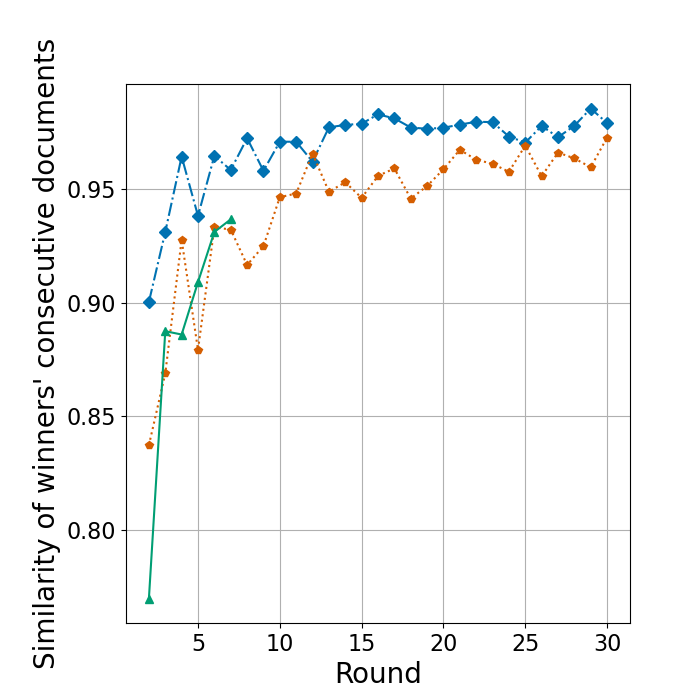}
        \subcaption{Class: M.}
        \label{label_rq1_winner_sim}
        \end{minipage}
        \hfill
        \begin{minipage}[t]{0.32\textwidth}
            \includegraphics[width=\linewidth]{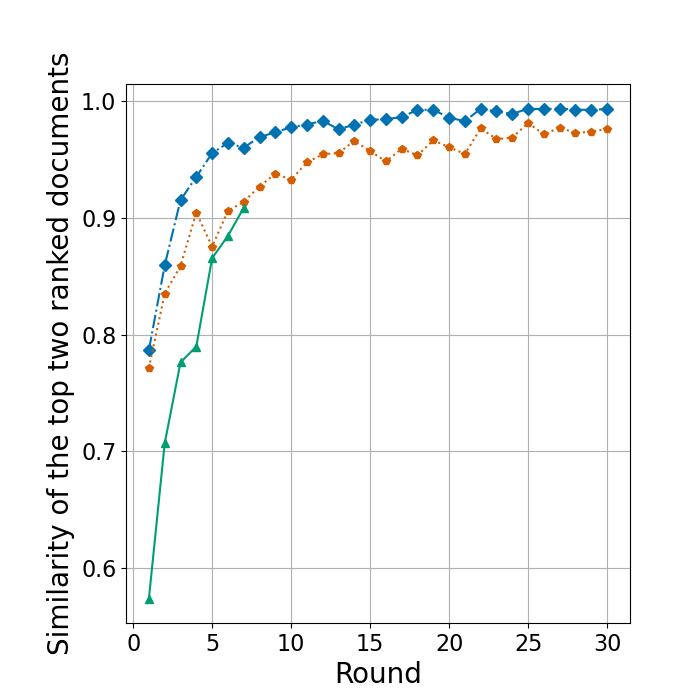}
            \subcaption{Class: M.}
            \label{label_rq1_top2_sim}
        \end{minipage}
        \hfill
        \begin{minipage}[t]{0.32\textwidth}
            \includegraphics[width=\linewidth]{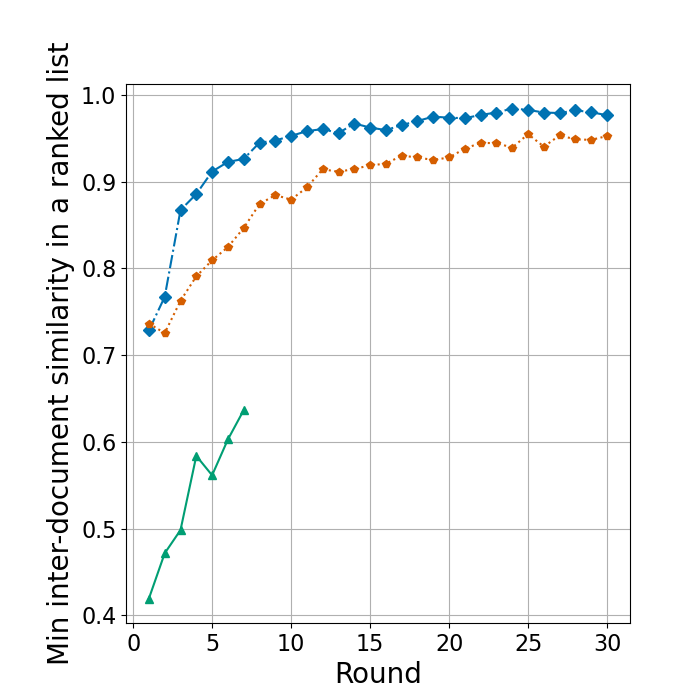}
            \subcaption{Class: D, C.}
            \label{label_rq1_group_diam}
        \end{minipage}
        \begin{minipage}[t]{0.32\textwidth}
            \includegraphics[width=\linewidth]{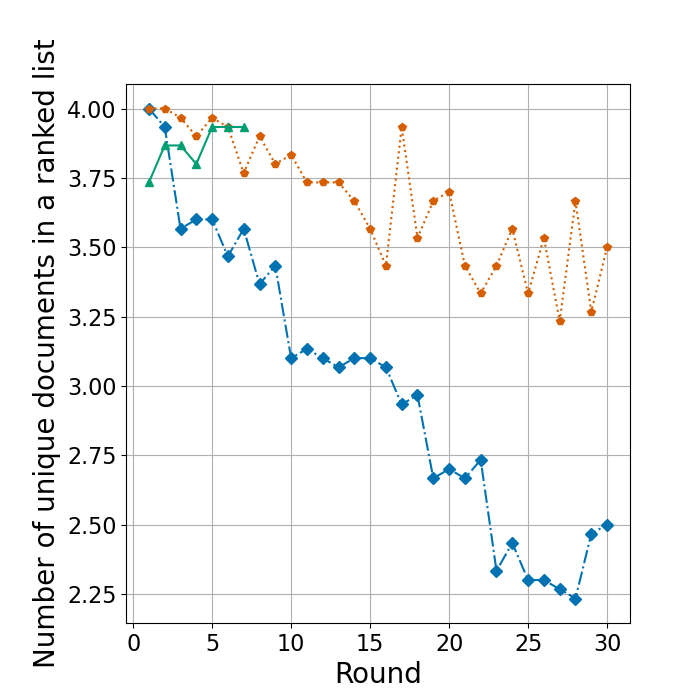}
            \subcaption{Class: D,C.}
            \label{label_rq1_unique}
        \end{minipage}
        \hfill
        \begin{minipage}[t]{0.32\textwidth}
            \centering
            \includegraphics[width=\linewidth]{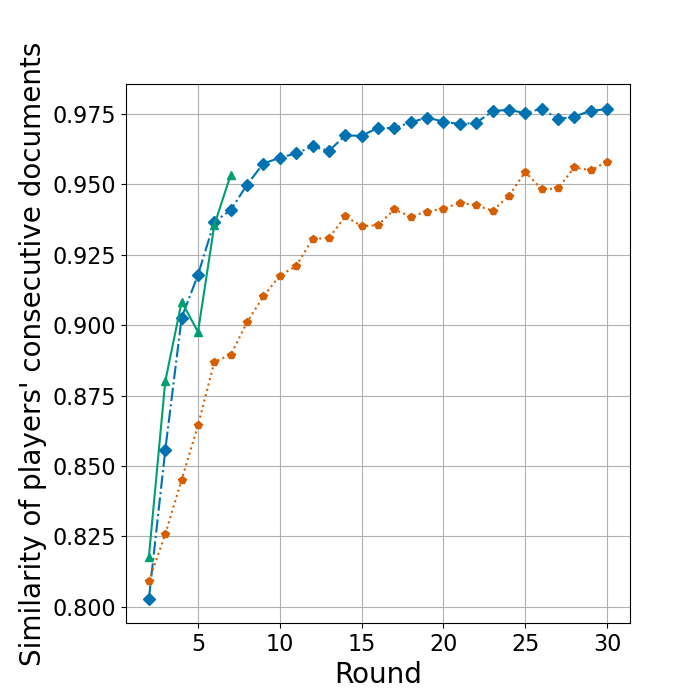}
            \subcaption{Class: D,C.}
            \label{label_rq1_conse_player}
        \end{minipage}
        \hfill
        % \begin{minipage}[c]{0.32\textwidth}
            % \includegraphics[width=\linewidth]{figs/RQ1/graphs_comparison/average_and_diameter_of_player_documents-min/bert_embeddings_similarity-min_player_documents.png}
            % \subcaption{(e) similarity metric: BERT; Class: Diversity, Convergence.}
                % \centering
            %     \includegraphics[width=\linewidth]{figs/RQ1/annotation_comparison/quality.png}
            %     \subcaption{Quality grades.}
            %     \label{label_rq1_quality}
            %     % \vspace{1pt}
            %     \includegraphics[width=\linewidth]{figs/RQ1/annotation_comparison/relevance.png}
            % \subcaption{Relevance grades.}
            % \label{label_rq1_relevance}
        % \end{minipage}
        \begin{minipage}[t]{0.32\textwidth}
            \centering
            \includegraphics[width=\linewidth]{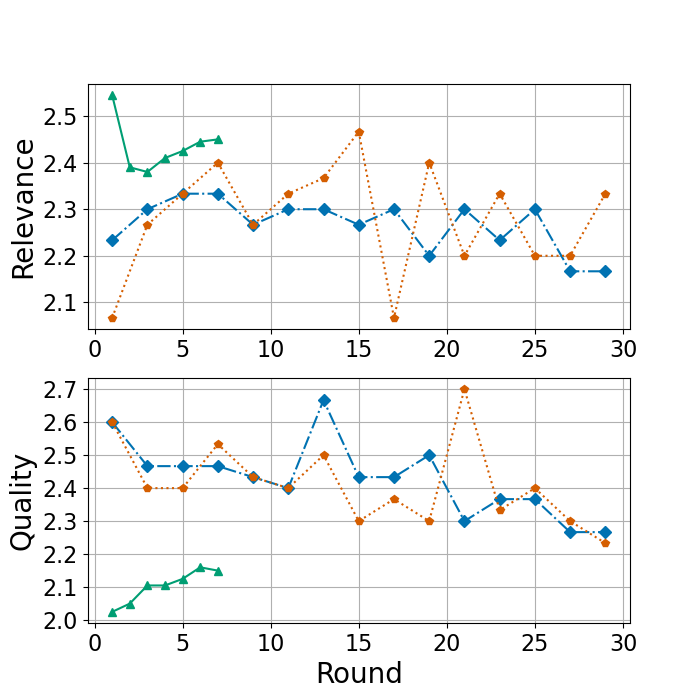}
            \subcaption{Class: R (top) and Q (bottom).}
            \label{label_rq1_annotation}
        \end{minipage}
    \end{minipage}
    \hfill
    \caption{Comparing LLM-based competitions with a competition with human players \cite{mordo_search_2025} (RQ1): {\LlamaEfiveListwiseFour}, {\LlamaEfivePairwiseFour} and {\ASRC} \cite{mordo_search_2025}. (a) average (over queries) similarity of winners between consecutive rounds ($i, i+1$); (b) the average (over queries) similarity between the two highest ranked documents; (c) the average (over queries) min inter-document similarity in a ranked list over rounds; (d) the number of unique documents over rounds; (e) the average (over queries and players) similarity of players' documents between consecutive rounds ($i, i+1$); (f) the average relevance (top) and quality (bottom) grades over rounds. \ASRC{} \cite{mordo_search_2025} consists of 15 games with seven rounds; The LLM-based competitions consist of 30 games with 30 rounds. We refer to the classes of measures \textit{\underline{M}imicking-the-winner}, \textit{\underline{D}iversity}, \textit{\underline{C}onvergence} and \textit{\underline{R}elevance and \underline{Q}uality} as \textbf{M}, \textbf{D}, \textbf{C}, \textbf{R} and \textbf{Q}, respectively.}
    \label{fig_compare_RQ1}
\end{figure}

Further exploration of the \textit{mimicking-the-winner} phenomenon is presented in Figure \ref{label_rq1_winner_sim}, where we analyze the similarity of winners' documents between consecutive rounds. For all three competitions, winners become increasingly similar across rounds. In the human competition, similarity levels are initially lower than the LLM-based competitions; however, they exhibit a steeper increase across rounds. These results are consistent with those presented in Figure \ref{fig_rq1_feature_strategies}, where a decreasing trend in feature value differences was observed. This decrease aligns with an increase in the similarity of winning documents across consecutive rounds.

In Figure \ref{label_rq1_top2_sim} we present the similarity between the top two ranked players' documents. The results indicate that in all three competitions, the top two ranked players become more similar as the competition progresses. Nevertheless, similarity levels remain slightly lower in the human competition. This observation may be attributed to the lower number of rounds in the human competition. Another possible explanation could be that LLM players have a similar configuration in competitions, leading them to generate highly similar documents. Recall that the only difference between LLM players in a competition was their persona.

We now turn to analyze the \textit{Diversity} class of measures. Figure \ref{label_rq1_group_diam} presents the minimum inter-document similarity in a ranked list over rounds. For all three competitions, as the game progresses, the documents in ranked lists become increasingly similar, leading to a decrease in ranked list diversity. Notably, the level of similarity in the two LLM-based competitions is higher than that observed in \ASRC{} \cite{mordo_search_2025}. Further analysis of the average number of unique documents (Figure \ref{label_rq1_unique}) reveals a general decreasing trend in LLM-based competitions, in contrast to a more stable trend in the human competition. Recall that non-unique documents means the players essentially copied other players' documents. By the end of seven rounds, the \ASRC{}'s dataset \cite{mordo_search_2025} contained an average of 3.8 unique documents per round. In comparison, a slightly lower number of unique documents was observed in the LLM-based competitions, with averages of 3.75 and 3.6 in the competitions involving Pairwise and Listwise agents, respectively.

% Listwise prompts, there is a sharper decline in the number of unique documents compared to the slower decline observed in the competition with Pairwise agents. In contrast, \ASRC{}'s human competition does not exhibit a clear trend in this regard. By the end of seven rounds, \ASRC{} \cite{mordo_search_2025} dataset included 3.8 unique documents (on average) per round, comparing a smaller amount of unique documents observed in the LLM competitions (3.75 and 3.6 in the competitions with the Pairwise and Listwise agents, respectively).

% Extending the analysis of the diversity in ranked lists to more rounds reveals that in the LLM competitions the diversity keep decreasing as the rounds advance. 

The findings just presented, along with the observed \textit{mimicking-the-winner} phenomenon (Figures \ref{fig_rq1_feature_strategies} and \ref{label_rq1_winner_sim}), attest that LLM-based agents tend to generate less diverse documents than human participants. This reduction in content diversity may lead to a sub-optimal user welfare \cite{basat_game_2017} and an increased herding effect, negatively impacting the topical diversity of the resulting corpus \cite{goren_driving_2021}. These effects appear to be more pronounced in LLM-based competitions compared to those involving human players.

Figure \ref{label_rq1_conse_player} illustrates the similarity of players' documents across consecutive rounds. At the beginning of all three competitions, players tended to make more substantial modifications to their documents. In \ASRC{} \cite{mordo_search_2025}, only seven rounds are available, limiting conclusions about longer-term trend of convergence and diversity. In both LLM-based competitions, the similarity of players' documents between consecutive rounds saturates and converges over time, providing empirical evidence that players modify their documents less as the rounds progress.

% Finally, \ref{label_rq1_} analyzes the similarity of consecutive documents produced by players. This analysis confirms that both human players and LLM-based players tend to modify their documents less as the game advances, demonstrating convergence at the individual player level as well.
% We emphasize the importance of using different similarity metrics since they give different insights about the dynamics of the game.

Figure \ref{label_rq1_annotation} shows that the quality and relevance grades are quite different between the LLMs and the human competitions. The quality grades in the human competition were lower than those in the LLM-based competitions, showing an increasing trend over the seven rounds, ranging from 2.05 to 2.15. In contrast, the quality grades in the LLM-based competitions ranged between 2.3 and 2.7, without a clear trend over the rounds. This finding aligns with previous reports indicating that LLMs consistently generate high-quality content in ranking competitions \cite{wu_survey_2024, nachimovsky_ranking-incentivized_2024}. Unlike the quality grades, relevance grades in the LLM competition were slightly lower compared to the human competition. In all three competitions, no clear trend in relevance grades over the rounds was observed.

We now turn to the analysis of the proportion of the wins. Recall from Section \ref{label_subsection_evaluation_measures} that it is computed as the ratio between the actual number of wins by the best-performing players and the expected number of wins, assuming that each player has an equal probability of winning a round (averaged over the games). In \ASRC{}'s dataset \cite{mordo_search_2025}, the resultant proportion of the wins was 2.62, compared to 1.4 observed in both LLM-based competitions. This suggests that in \ASRC{} \cite{mordo_search_2025}, there is often a dominant player, whereas in LLM-based competitions, wins are more evenly shared among players. One possible explanation for this difference is the varying levels of motivation among human players, as individual participants may have different incentives that influence their engagement and performance. Another explanation for the reduced proportion of the wins in the LLM-based competitions is that the LLM-based players have similar characteristics, which leads them to exhibit more uniform behavior.

\newcommand{\figWidth}{1.1in}
\newcommand{\figHeight}{1.1in}
\newcommand{\redSpace}{-.1in}
\begin{figure}[t]
\centering 
\includegraphics[scale=0.35]{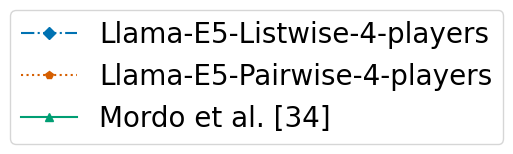}
  \begin{tabular}{ccc}
    \multicolumn{3}{c}{\textbf{Query dependent features}} \\
    \includegraphics[width=\figWidth,height=\figHeight]{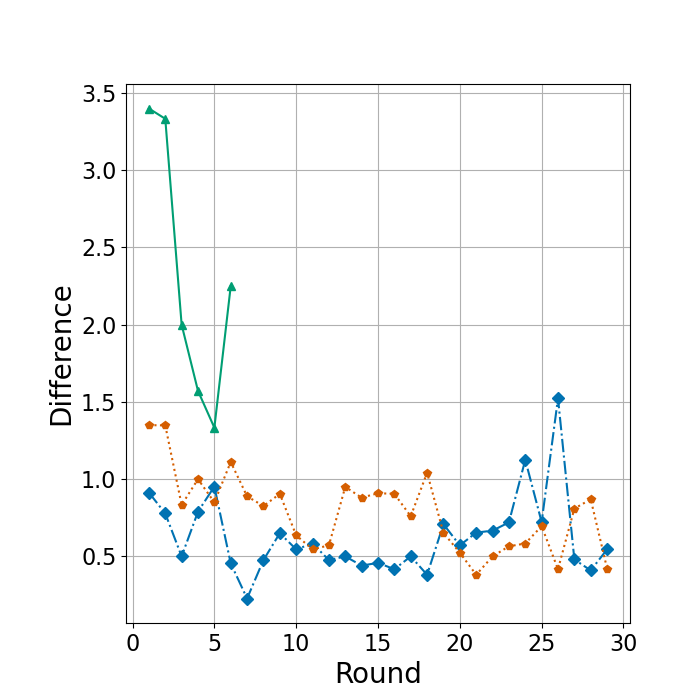} & \hspace*{\redSpace} \includegraphics[width=\figWidth,height=\figHeight]{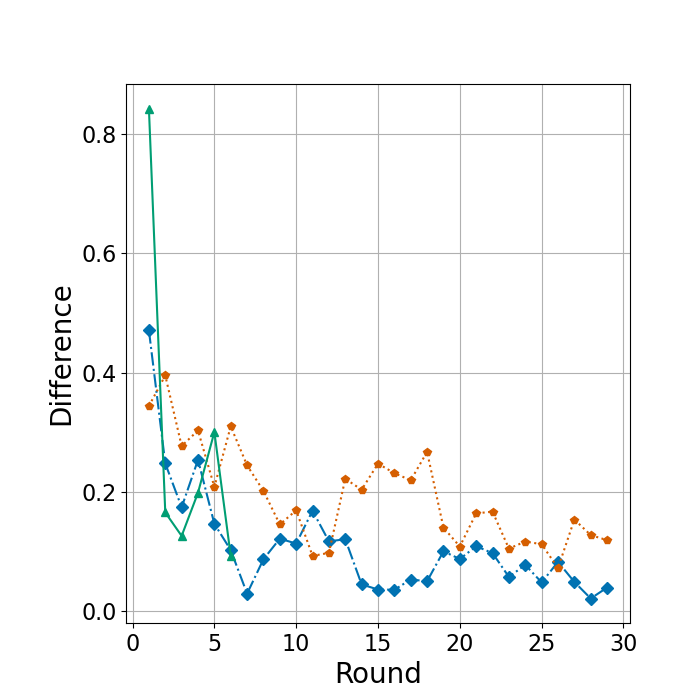} & \hspace*{\redSpace} \includegraphics[width=\figWidth,height=\figHeight]{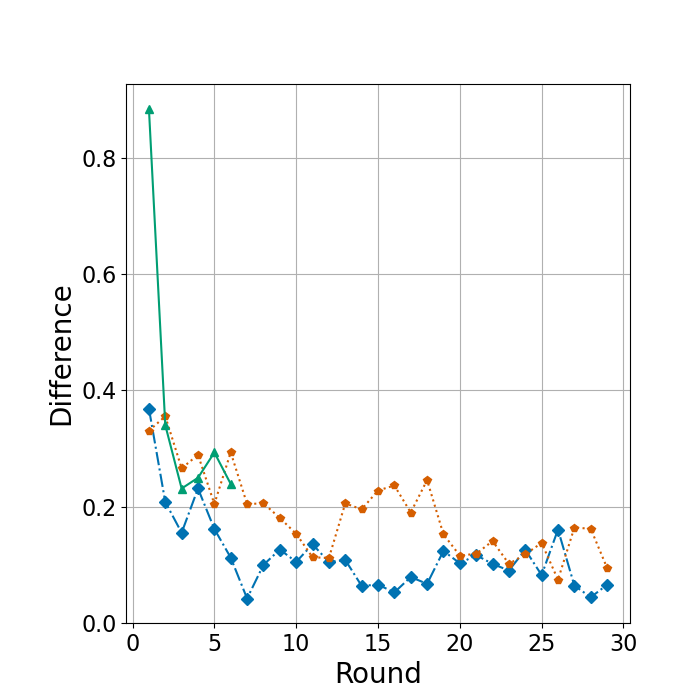} \\
    TF & BM25 & LM.DIR \\
    \multicolumn{3}{c}{\textbf{Query independent features}} \\
    \includegraphics[width=\figWidth,height=\figHeight]{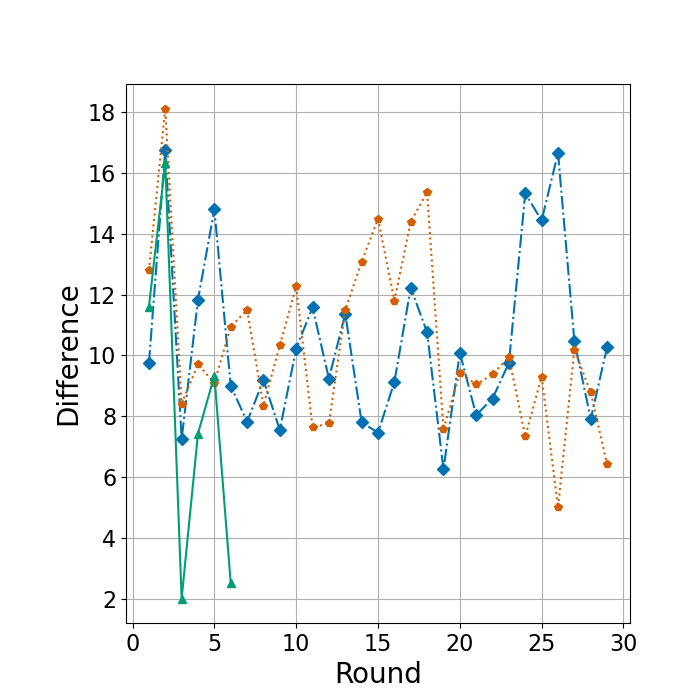} & \hspace*{\redSpace} \includegraphics[width=\figWidth,height=\figHeight]{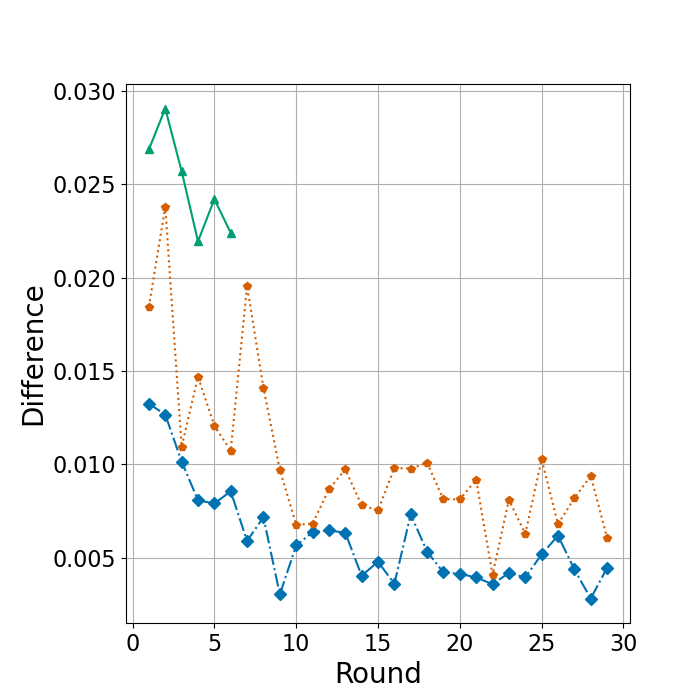} & \hspace*{\redSpace} \includegraphics[width=\figWidth,height=\figHeight]{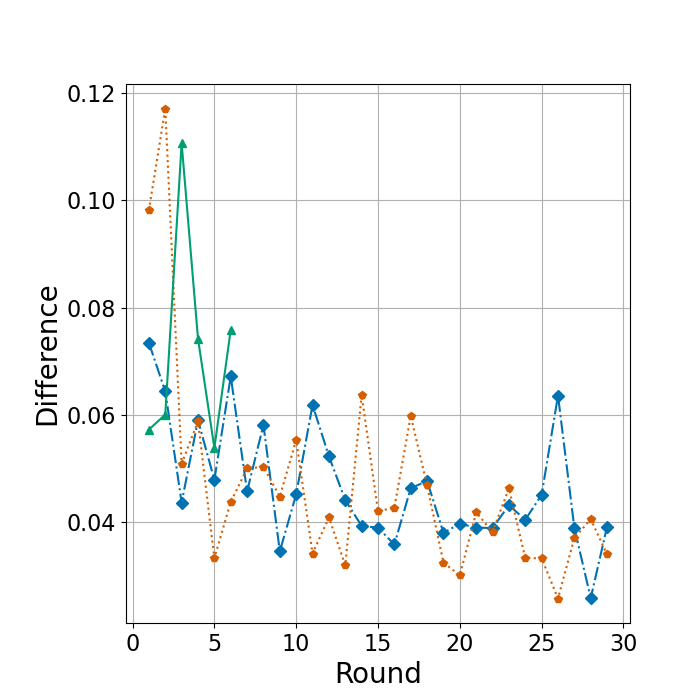} \\
    Length & StopwordRatio & Entropy \\
  \end{tabular}
  \caption{\label{fig_rq1_feature_strategies} Average absolute difference of feature values of winner documents in rounds $i$ ($W_i$) and $i+1$ ($W_{i+1}$).}
\end{figure}

\subsection{Results (RQ2): comparing LLM-based agents}\label{section_rq2}
We proceed to compare different LLM-based competitions while keeping the ranking function fixed as the cosine of document and query E5 embeddings, henceforth referred to as the E5 ranking function. To potentially increase the competition dynamics with respect to the four-players competitions that were addressed in RQ1, we increased the number of players in each game to five. We explored variations of the competition with the Listwise prompt\footnote{Recall that Bardas et al. \cite{bardas_prompt-based_2025} identified it as one of the two best-performing \contextualized{} of the prompt.} (\LlamaEfiveListwise{}): (1) changing the \contextualized{} to Pairwise (\LlamaEfivePairwise{}), (2) modifying the \shared{} to instruct the LLM not to copying other players' documents (\LlamaEfiveListwiseNoCopy{}), and (3) replacing Llama with Gemma as the LLM (\GemmaEfiveListwise{}).

% The full list of parameters of every competition appears in Appendix \ref{appendix_dataset_parameterts} Tables \ref{table_data_param} and \ref{table_datasets}.

\begin{figure}[ht!]
    \centering
    \includegraphics[scale=0.35]{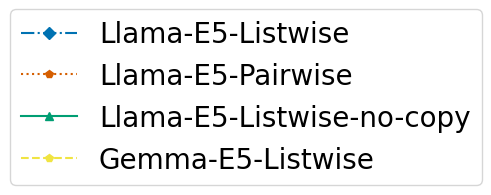}
    \begin{minipage}[t]{0.48\textwidth}
        \centering
        \begin{minipage}[t]{0.32\textwidth}
            \includegraphics[width=\linewidth]{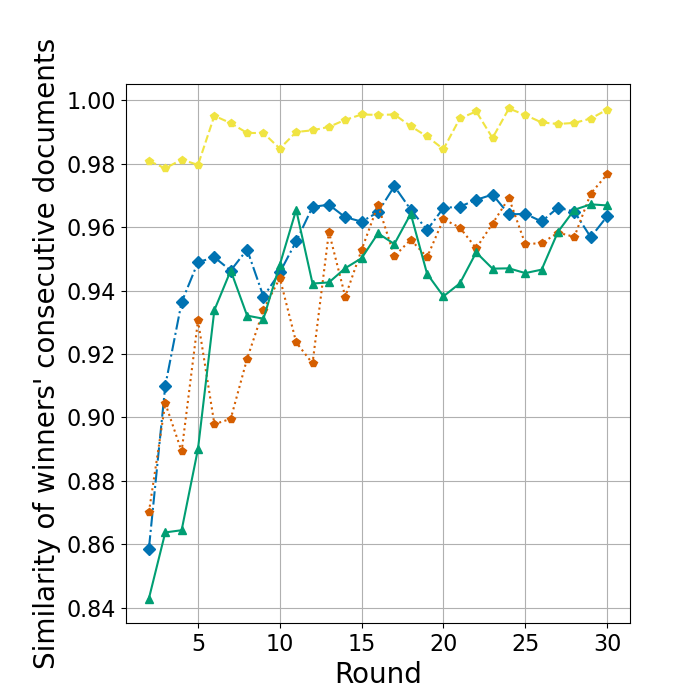}
        \subcaption{Class: M.}
        \label{label_rq2_winner_sim}
        \end{minipage}
        \hfill
        \begin{minipage}[t]{0.32\textwidth}
            \includegraphics[width=\linewidth]{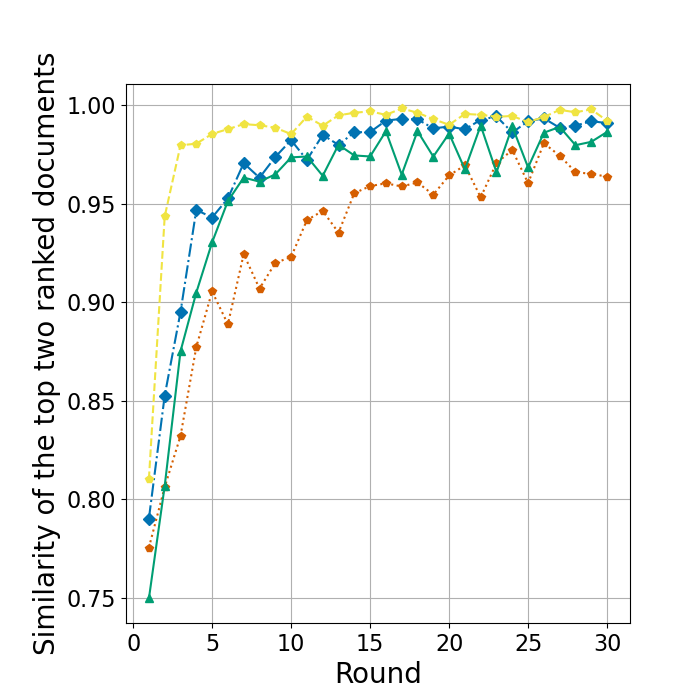}
            \subcaption{Class: M.}
            \label{label_rq2_top2_sim}
        \end{minipage}
        \hfill
        \begin{minipage}[t]{0.32\textwidth}
            \includegraphics[width=\linewidth]{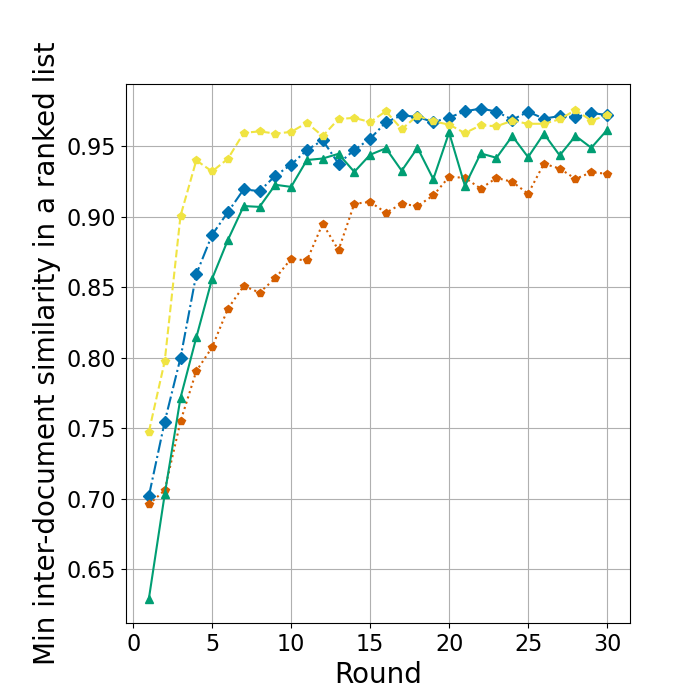}
            \subcaption{Class: D, C.}
            \label{label_rq2_group_diam}
        \end{minipage}        
        \begin{minipage}[t]{0.32\textwidth}
            \includegraphics[width=\linewidth]{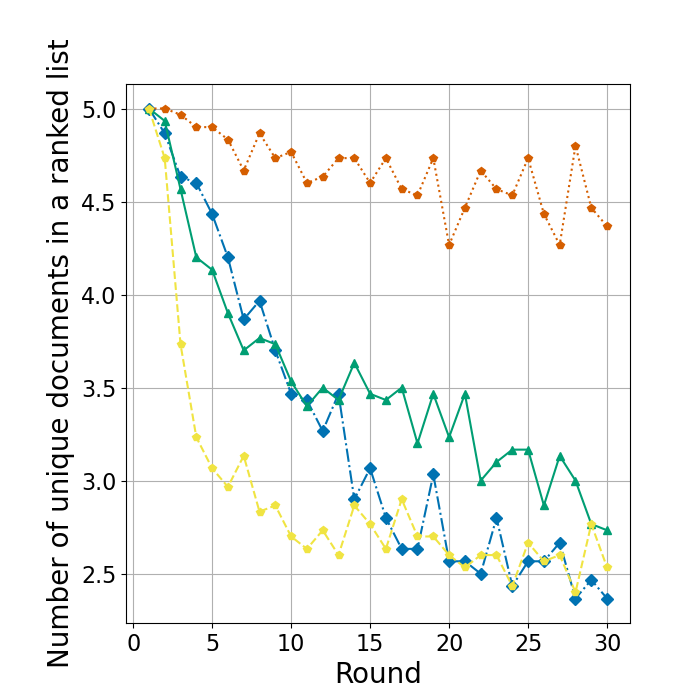}
            \subcaption{Class: D, C.}
            \label{label_rq2_unique}
        \end{minipage}
        \hfill
        \begin{minipage}[t]{0.32\textwidth}
            \includegraphics[width=\linewidth]{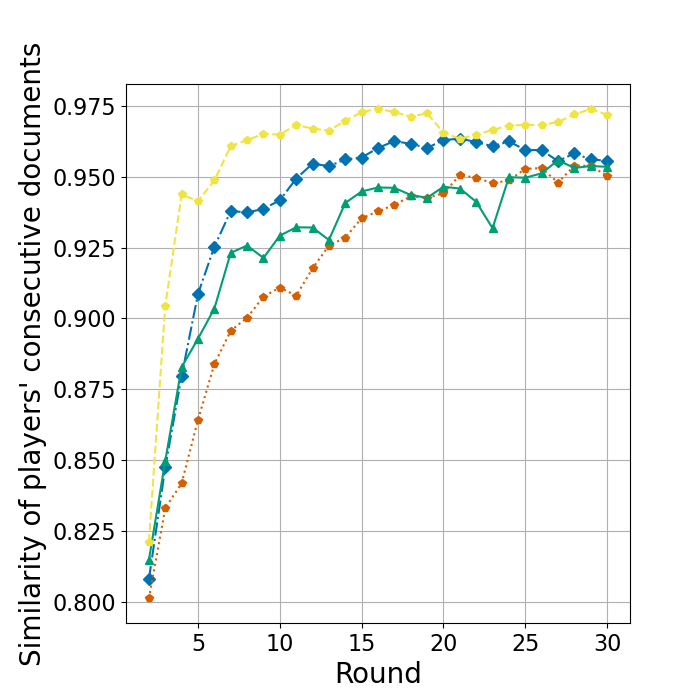}
            \subcaption{Class: D, C.}
            \label{label_rq2_conse_player}
        \end{minipage}
        \hfill
        % \begin{minipage}[c]{0.32\textwidth}
            % \includegraphics[width=\linewidth]{figs/RQ1/graphs_comparison/average_and_diameter_of_player_documents-min/bert_embeddings_similarity-min_player_documents.png}
            % \subcaption{(e) similarity metric: BERT; Class: Diversity, Convergence.}
        %         \centering
        %         \includegraphics[width=\linewidth]{figs/RQ2/annotation_comparison/quality.png}
        %         \subcaption{Quality grades.}
        %         \label{label_rq2_quality}
        %         % \vspace{1pt}
        %         \includegraphics[width=\linewidth]{figs/RQ2/annotation_comparison/relevance.png}
        %     \subcaption{Relevance grades.}
        %     \label{label_rq2_relevance}
        % \end{minipage}
        \begin{minipage}[t]{0.32\textwidth}
            \centering
            \includegraphics[width=\linewidth]{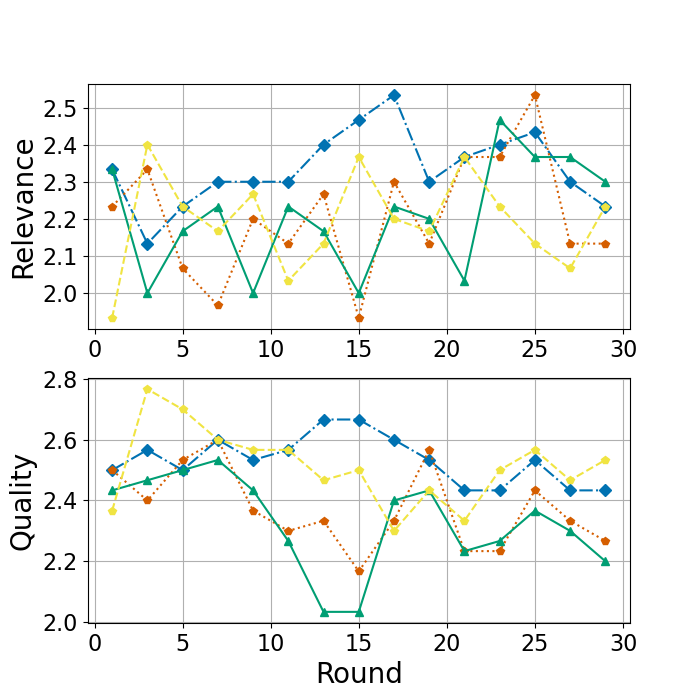}
            \subcaption{Class: R (top) and Q (bottom).}
            \label{label_rq2_annotation}
        \end{minipage}
        \hfill
    \end{minipage}
    \hfill
    \caption{Comparison of LLM-based agents (RQ2): \LlamaEfiveListwise{}, {\LlamaEfivePairwise{}}, {\LlamaEfiveListwiseNoCopy} and {\GemmaEfiveListwise{}}. (a) average (over queries) similarity of winners between consecutive rounds ($i, i+1$); (b) the average (over queries) similarity between the two highest ranked documents; (c) the average (over queries) min inter-document similarity in a ranked list over rounds; (d) the number of unique documents over rounds; (e) the average (over queries and players) similarity of players' documents between consecutive rounds ($i, i+1$); (f) the average relevance (top) and quality (bottom) grades over rounds. Each competition consists of 30 games with 30 rounds. We refer to the classes of measures \textit{\underline{M}imicking-the-winner}, \textit{\underline{D}iversity}, \textit{\underline{C}onvergence} and \textit{\underline{R}elevance and \underline{Q}uality} as \textbf{M}, \textbf{D}, \textbf{C}, \textbf{R} and \textbf{Q}, respectively.}
    \label{fig_compare_RQ2}
\end{figure}

In the \textit{Mimicking-the-winner} class of measures, both the similarity of winners' documents between consecutive rounds (Figure \ref{label_rq2_winner_sim}) and the similarity between the top two players (Figure \ref{label_rq2_top2_sim}) generally increased over round across all competitions. Notably, the similarity levels for Gemma were consistently higher than those for the other competitions and also leveled off way more quickly. Overall, the Gemma-LLM exhibits a stronger tendency to mimic documents compared to the Llama LLM competitions.

The minimum inter-document similarity over rounds (Figure \ref{label_rq2_group_diam}) is higher in the competition with Gemma-LLM compared to all other three competitions with Llama-based agents. It attests to a lower ranked list diversity in the Gemma-based competition compared to the Llama-based competitions. The analysis of unique documents (Figure \ref{label_rq2_unique}) reveals a decline in the number of unique documents across rounds in all competitions. Notably, the competition involving Gemma players exhibits the steepest decrease, with the proportion of unique documents converging to approximately 50\%. This results in an average of 2.5 unique documents out of a total of 5 in the ranked list. In the competition with the Pairwise prompts, the copying behavior was observed to a smaller extent. Between the two Listwise competitions, the \LlamaEfiveListwiseNoCopy{} competition included more unique documents on average, consistent with the instructions in the \shared{} to avoid copying documents of other players.

As for the \textit{Convergence} class, examination of the similarity of players' documents between consecutive rounds (Figure \ref{label_rq2_conse_player}) reveals that the competition with Gemma-LLM (\GemmaEfiveListwise{}) exhibits an increasing and higher levels of similarity over rounds, in comparison to the Llama-LLM competitions (\LlamaEfiveListwise{}, \LlamaEfivePairwise{} and \LlamaEfiveListwiseNoCopy{}). We conclude that the competition with Gemma-LLM converges more rapidly comparing to the other competitions.
Overall, our findings suggest that the most influential factor that affects the dynamics of the competitions, as reflected by the measures, is the language model of the agents rather than the provided prompts. As for the Llama competitions, a slightly higher levels of \textit{mimicking-the-winner} were observed, with respect to the corresponding measures, when the Listwise prompt was used compared to the Pairwise prompt. This observation may be attributed to structural differences between the prompts: the Pairwise prompt provides a limited context regarding other rankings, whereas the Listwise prompt contains a broader view of competing players. Consequently, this broader context facilitates the adoption of the strategy proven to be optimal \cite{raifer_information_2017}, which involves mimicking winning documents.

The quality grades (Figure \ref{label_rq2_annotation}) in the LLM competitions show a marginal tendency to decrease over rounds. This suggests that the competition dynamics may contribute to a gradual decline in overall corpus quality. Moreover, competitions with the Listwise prompts, specifically \LlamaEfiveListwise{} and \GemmaEfiveListwise{}, included documents with slightly higher quality compared to \LlamaEfivePairwise{} and \LlamaEfiveListwiseNoCopy{}. As for the relevance grades (Figure  \ref{label_rq2_annotation}), the \LlamaEfiveListwise{} competition exhibits marginally higher relevance grades across most rounds compared to other competitions. Specifically, the competition using prompts that explicitly instruct players to avoid copying (\LlamaEfiveListwiseNoCopy{}) appears to yield lower quality and relevance grades across most rounds compared to the competition without such an instruction (\LlamaEfiveListwise{}). A possible explanation is that restricting copying may hinder players from adopting the \textit{mimicking-the-winner} strategy \cite{raifer_information_2017}, which is associated with achieving higher rankings. Since top-ranked documents are generally more relevant, this restriction could negatively impact overall document quality and relevance. 

\subsection{Results (RQ3): Llama agents with different rankers
}\label{section_rq3}

We next turn to study the effect of the ranking function on the competition. To this end, we fix the LLM to Llama3.1, and study the ranking functions: the cosine of document and
query E5 \cite{wang_text_2024} embeddings (refer to as E5 ranking function), the cosine of document and
query Contriever embeddings \cite{izacard_unsupervised_2022} (refer to as Contriever ranking function), and Okapi \cite{maron_relevance_1960}. We used agents with both Pairwise and Listwise prompts.

\begin{figure}[ht!]
    \centering
    \includegraphics[scale=0.35]{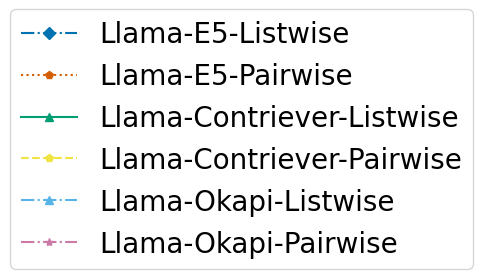}
    \begin{minipage}[t]{0.47\textwidth}
        \centering
        \begin{minipage}[t]{0.32\textwidth}
            \includegraphics[width=\linewidth]{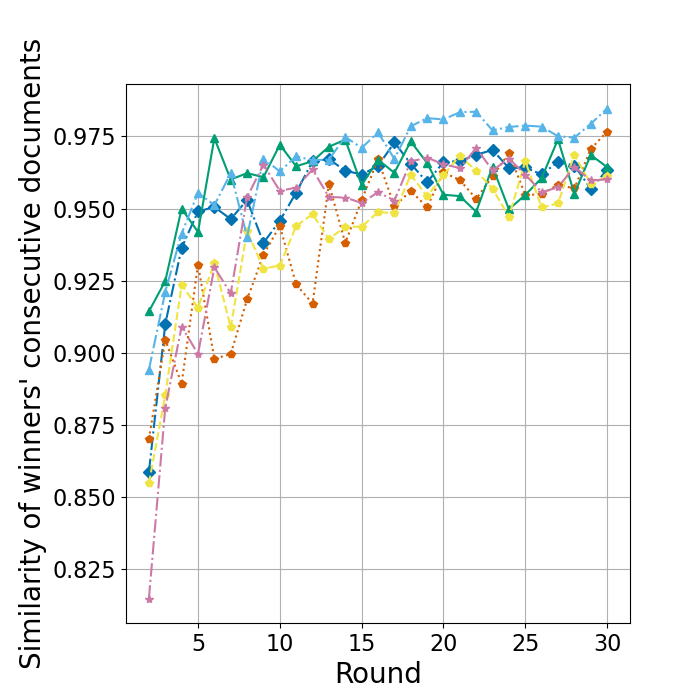}
        \subcaption{Class: M.}
        \label{label_rq3_winner_sim}
        \end{minipage}
        \hfill
        \begin{minipage}[t]{0.32\textwidth}
            \includegraphics[width=\linewidth]{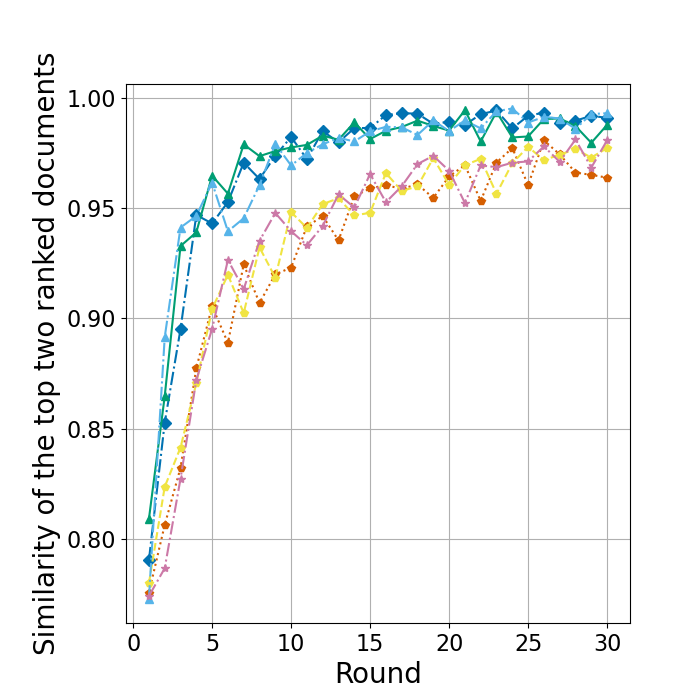}
            \subcaption{Class: M.}
            \label{label_rq3_top2_sim}
        \end{minipage}
        \hfill
        \begin{minipage}[t]{0.32\textwidth}
            \includegraphics[width=\linewidth]{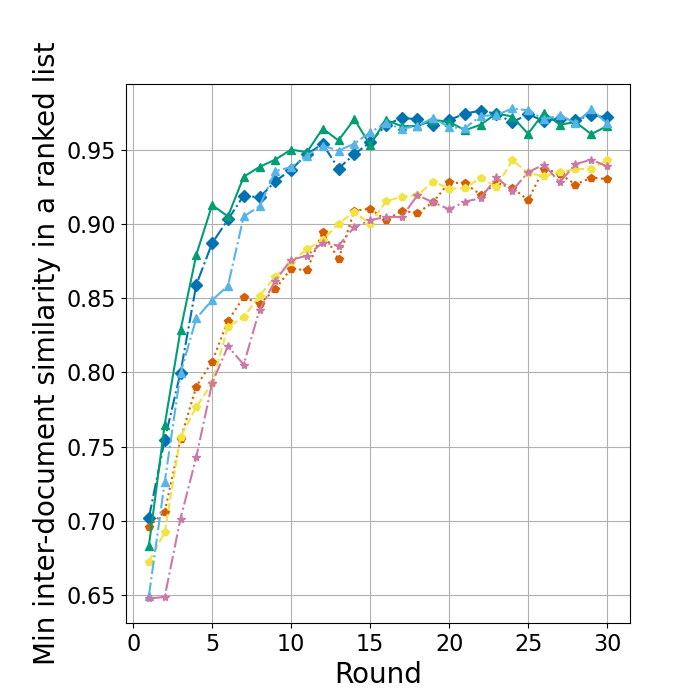}
            \subcaption{Class: D, C.}
            \label{label_rq3_group_diam}
        \end{minipage}
        % Stop Cover
        \begin{minipage}[t]{0.32\textwidth}
            \includegraphics[width=\linewidth]{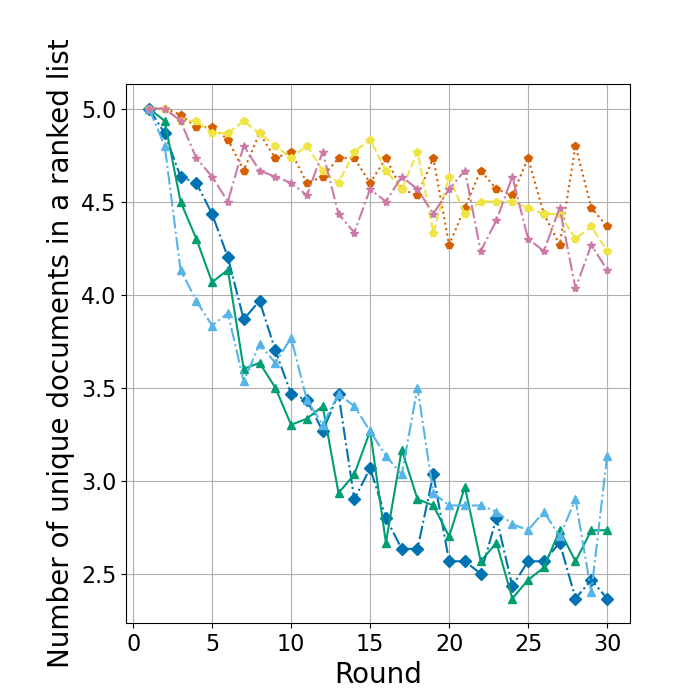}
            \subcaption{Class: D, C.}
            \label{label_rq3_unique}
        \end{minipage}
        \hfill
        \begin{minipage}[t]{0.32\textwidth}
            \includegraphics[width=\linewidth]{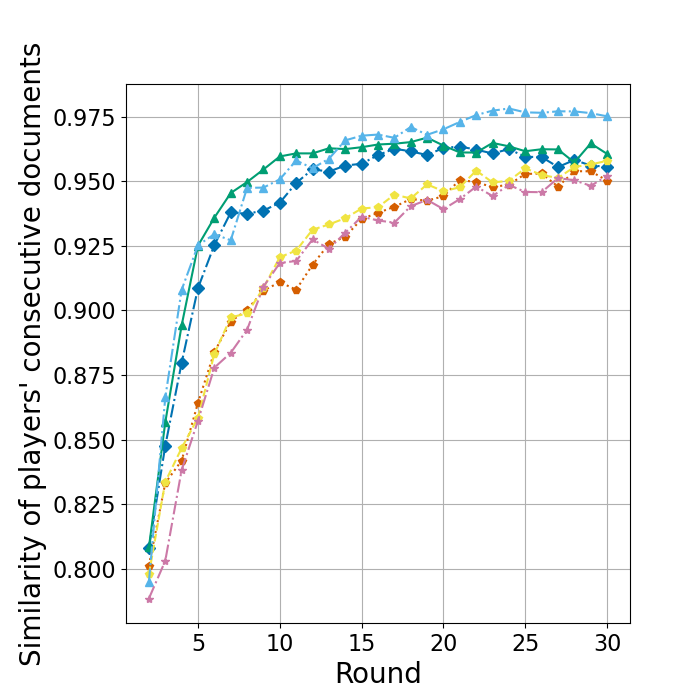}
            \subcaption{Class: D, C.}
            \label{label_rq3_conse_player}
        \end{minipage}
        \hfill
        % \begin{minipage}[c]{0.32\textwidth}
            % \includegraphics[width=\linewidth]{figs/RQ1/graphs_comparison/average_and_diameter_of_player_documents-min/bert_embeddings_similarity-min_player_documents.png}
            % \subcaption{(e) similarity metric: BERT; Class: Diversity, Convergence.}
        %         \centering
        %         \includegraphics[width=\linewidth]{figs/RQ3/annotation_comparison/quality.png}
        %         \subcaption{Quality grades.}
        %         \label{label_rq3_quality}
        %         % \vspace{1pt}
        %         \includegraphics[width=\linewidth]{figs/RQ3/annotation_comparison/relevance.png}
        %         \subcaption{(g) Relevance grades.}
        %         \label{label_rq3_relevance}
        % \end{minipage}
        \begin{minipage}[t]{0.32\textwidth}
            \includegraphics[width=\linewidth]{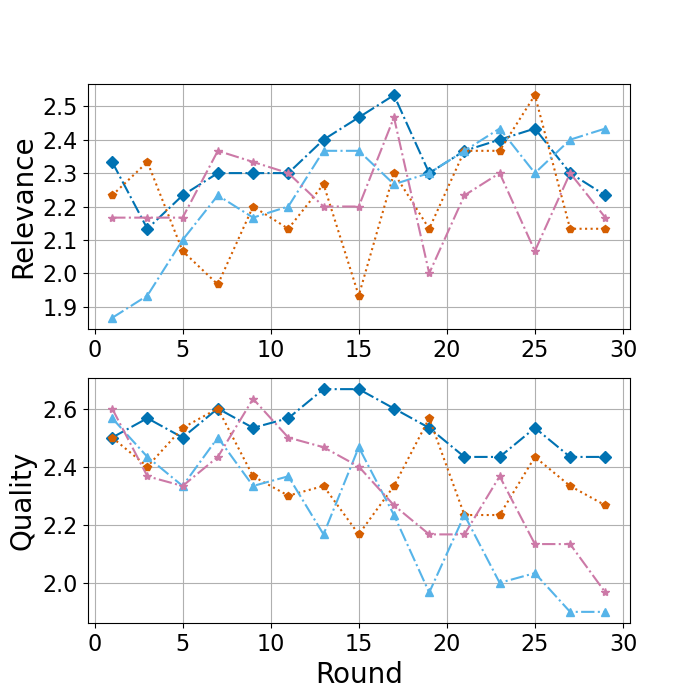}
            \subcaption{Class: R (top) and Q (bottom).}
            \label{label_rq3_annotation}
        \end{minipage}
    \end{minipage}
    \hfill
    \caption{Comparison of ranking functions in LLM-based competitions (RQ3): \LlamaEfiveListwise{}, \LlamaEfivePairwise{}, \LlamaContListwise{}, \LlamaContPairwise{}, \LlamaOkapiListwise{}, and \LlamaOkapiPairwise{}. (a) average (over queries) similarity of winners between consecutive rounds ($i, i+1$); (b) the average (over queries) similarity between the two highest ranked documents; (c) the average (over queries) min inter-document similarity in a ranked list over rounds; (d) number of unique documents over rounds; (e) the average (over queries and players) similarity of players' documents between consecutive rounds ($i, i+1$); (f) the average relevance (top) and quality (bottom) grades over rounds. Each competition consists of 30 games with 30 rounds. We refer to the classes of measures \textit{\underline{M}imicking-the-winner}, \textit{\underline{D}iversity}, \textit{\underline{C}onvergence} and \textit{\underline{R}elevance and \underline{Q}uality} as \textbf{M}, \textbf{D}, \textbf{C}, \textbf{R} and \textbf{Q}, respectively.}
    \label{fig_compare_RQ3}
\end{figure}

The average similarity between winners (Figure \ref{label_rq3_winner_sim}) was similar across the ranking functions. At the end of the competition with the Okapi ranking function and players with the Listwise prompt, an higher similarity with respect to consecutive winners was observed, compared to other competitions. Additionally, the average similarity between the top two players (Figure \ref{label_rq3_top2_sim}) increased over rounds. For all ranking functions, the choice of ranking function did not substantially affect both measures of the \textit{Mimicking-the-winner} class. As discussed in Section \ref{section_rq2} and illustrated in Figure \ref{label_rq3_top2_sim}, the type of prompt (Listwise vs. Pairwise) has a significantly greater impact than the ranking function.

As for the diversity at the ranked-list-level, measured by both the minimum inter-document similarity in the ranked list and the number of unique documents, we observe the same trend of being unaffected by the ranking function (Figures \ref{label_rq3_group_diam} and \ref{label_rq3_unique}). We can see two distinct "clusters" for both the Listwise and Pairwise LLM-based competitions, with minimal impact with respect to the choice of the ranking function. The convergence and diversity at the player level measured by the similarity of players' documents between consecutive rounds (Figure \ref{label_rq3_conse_player}), demonstrated an increasing trend (over rounds) across all the competitions. Once again, the differences observed across ranking functions remain minimal. However, in the competition with the Okapi ranking function and agents with the Listwise prompt, slightly elevated similarity levels were observed in the final rounds compared to all other competitions.

The quality and relevance grades were similar across all the competitions. (See Figure \ref{label_rq3_annotation}.) The quality grades across all competitions (excluding those with Contriever ranking functions, which were not annotated) exhibit a general tendency to decline over the rounds. Interestingly, the quality grades of \LlamaEfiveListwise{} were slightly higher compared to all other competitions. For relevance grades, the values consistently exceed 1.9 (out of possible 3) in most rounds, with no significant influence observed by the choice of ranking function.

Overall, the ranking function appears to have minimal to no significant impact on the competition dynamics with respect to the analyzed measures. We showed above that the type of LLM and the prompt have a much larger effect.

% The difference between listwise and pairwise prompts is further evident in group-level diversity, as indicated by the group diameter similarity (measured using SBERT) and the average similarity between consecutive documents (measured using TF.IDF). Using both lexical and semantic similarity metrics, we found that listwise prompts consistently resulted in lower diversity levels compared to pairwise prompts across all three ranking functions. This conclusion is also supported by the analysis of unique documents, which showed that competitions using listwise prompts produced fewer unique documents than those using pairwise prompts, regardless of the ranking function.

% The only metric where the ranking function noticeably influenced competition dynamics was the diameter similarity of individual players' documents when similarity was measured using BERT. While other similarity metrics produced slightly different results, the distinction between listwise and pairwise prompts remained evident across all ranking functions.\ref{label_rq1_winner_sim}

\section{Conclusions and Future Work}
We introduced a novel simulation platform, the \platformName{} \platform, designed to simulate ranking competitions with LLMs as document authors. Along with the platform, we presented the \platformName{} \analyzer and \platformName{} \compare tools, which facilitate the analysis of individual competitions and the comparison of multiple competitions. We demonstrated the capabilities of \platformName{} by generating multiple datasets with LLM-based agents and comparing the results across different datasets. Additionally, we compared LLM-based competitions with a human-based competition to study the behavioral dynamics of the two types of agents 
(LLM and human).

We found that LLMs exhibit strategic behavior similar to that of human players \cite{basat_game_2017, raifer_information_2017}, with an even greater tendency to adopt the \textit{mimicking-the-winner} strategy. This strategy results in herding behavior towards the top-ranked players \cite{goren_driving_2021}, leading to a reduced diversity in the corpus \cite{mordo_search_2025}. Further exploration revealed that the most influential factor affecting the behavior of LLMs is the language model they use, followed by the prompt provided. In contrast, the choice of ranking function had minimal to no effect on the dynamics of the competition.

For future work, we plan to use \platformName{} so as to study competitions that include different types of LLMs competing against each other, as well as humans competing against LLM agents.

% \begin{itemize}
%     \item Online (with human games)
%     \item multiple queries
%     \item mix llms-humans
%     \item planted documents
%     \item players are familiar with the number of round
%     \item 
% \end{itemize}
% XXX XXX XXX XXX XXX XXX XXX XXX XXX XXX XXX XXX XXX XXX XXX XXX XXX XXX XXX XXX XXX XXX XXX XXX XXX XXX XXX XXX XXX XXX XXX XXX XXX XXX XXX XXX XXX XXX XXX XXX XXX XXX XXX XXX XXX XXX XXX XXX XXX XXX XXX XXX XXX XXX XXX XXX XXX XXX XXX XXX 
% \input{appendix}
\balance
\bibliographystyle{ACM-Reference-Format}
% \bibliography{bib_simulation}

\end{document}